%% file: em_manuscript.tex
\documentclass[preprint,10pt]{elsarticle}
\pdfoutput=1


\makeatletter
\def\ps@pprintTitle{%
   \let\@oddhead\@empty
   \let\@evenhead\@empty
   \def\@oddfoot{\reset@font\hfil\thepage\hfil}
   \let\@evenfoot\@oddfoot
}
\makeatother

\input{packages.tex}
\usepackage[margin=1in]{geometry}
\biboptions{sort&compress}

\begin{document}

\begin{frontmatter}
\title{Characterization and Integration of the Singular Test Integrals in the Method-of-Moments Implementation of the Electric-Field Integral Equation}

\author[freno]{Brian A.\ Freno}
\ead{bafreno@sandia.gov}
\author[freno]{William A.\ Johnson}
\author[freno]{Brian F.\ Zinser}
\author[wilton]{Donald R. Wilton}
\author[vipiana]{Francesca Vipiana}
\author[freno]{Salvatore Campione}

\address[freno]{Sandia National Laboratories, Albuquerque, NM 87185}
\address[wilton]{University of Houston, Houston, TX 77204}
\address[vipiana]{Politecnico di Torino, 10129 Torino, Italy}

\begin{abstract}
\input{abstract.tex}
\end{abstract}

\begin{keyword}
method of moments \sep
singular integrals \sep
geometrically symmetric quadrature rules
\end{keyword}

\end{frontmatter}

\input{introduction.tex}

\input{singularity.tex}
\input{quadrature_rules.tex}
\input{results.tex}
\input{conclusions.tex}
\input{acknowledgments.tex}
\appendix
\renewcommand*{\thesection}{Appendix \Alph{section}}
\renewcommand*{\thesubsection}{\Alph{section}.\arabic{subsection}}
\setcounter{figure}{0}
\setcounter{table}{0}
\input{appendix_a.tex}
\setcounter{figure}{0}
\setcounter{table}{0}
\input{appendix_b.tex}

\addcontentsline{toc}{section}{\refname}
\bibliographystyle{elsarticle-num}
\bibliography{../quadrature_manuscript/quadrature.bib}

\end{document}

%% file: abstract.tex
In this paper, we characterize the logarithmic singularities arising in the method of moments from the Green's function in integrals over the test domain, and we use two approaches for designing geometrically symmetric quadrature rules to integrate these singular integrands.
These rules exhibit better convergence properties than quadrature rules for polynomials and, in general, lead to better accuracy with a lower number of quadrature points.  We demonstrate their effectiveness for several examples encountered in both the scalar and vector potentials of the electric-field integral equation (singular, near-singular, and far interactions) as compared to the commonly employed polynomial scheme and the double Ma--Rokhlin--Wandzura (DMRW) rules, whose sample points are located asymmetrically within triangles.

%% file: introduction.tex
\section{Introduction}
\label{sec:introduction}
The method of moments (MoM) is a useful technique in computational electromagnetics for solving the electric-field integral equation (EFIE), the magnetic-field integral equation (MFIE), and the combined-field integral equation (CFIE), upon discretizing surfaces using planar or curvilinear mesh elements.  Through this approach, four-dimensional integrals are evaluated, which integrate over source and test elements.  However, the presence of a Green's function in these equations yields scalar and vector potential terms with singularities (in their higher-order derivatives) when the test and source elements share one or more edges or vertices and near-singularities when they are otherwise close.

Many approaches have been developed to address the singularity and near-singularity for the inner, source-element integral.  While, originally, singularity subtraction schemes were proposed~\cite{graglia_1993,wilton_1984,rao_1982}, more recent approaches use singularity cancellation schemes~\cite{khayat_2005,fink_2008,khayat_2008,vipiana_2011,vipiana_2012,botha_2013}, through which the Jacobian from a variable transformation cancels the (near-)singularity, permitting the use of Gauss--Legendre quadrature rules.  More recently, a hybrid scheme that combines these two methods has been proposed~\cite{rivero_2019}.

Approaches have also been developed to address the singularity in the outer, test-element integral.  In~\cite{vipiana_2013}, the authors use the outer product of one-dimensional rules from~\cite{ma_1996}, which they map to the triangular test element through a Duffy transformation~\cite{duffy_1982}.  
In~\cite{polimeridis_2013}, the authors use a series of variable transformations and integration reordering to integrate the four-dimensional integrals. 
In~\cite{wilton_2017}, the authors present an approach for coplanar source and test elements, extended in~\cite{rivero_2019b} to general element orientations.
In~\cite{ylaoijala_2003}, for the CFIE, the authors avoid the singularity in the test-element integral by modifying the integrand.  In~\cite{gurel_2005}, for the MFIE, the authors use a singularity-extraction method for the test-element integral, which they use in~\cite{ergul_2005} to implement the MFIE in a manner that eliminates some of the restrictions due to the singularity.  In~\cite{polimeridis_2012}, the authors use double-exponential quadrature integration schemes.  In~\cite{tihon_2018}, the authors expand the integrand in a (truncated) power series and analytically integrate term by term.  This approach, however, cannot be applied to integrals that do not involve a homogeneous-medium Green's function. 

In this paper, we characterize the logarithmic singularities in the test integral and use geometrically symmetric quadrature rules better suited for evaluating them.  We compare our rules with a standard polynomial scheme~\cite{dunavant_1985} and with the asymmetric double Ma--Rokhlin--Wandzura (DMRW) rules~\cite{vipiana_2013,ma_1996}.  In particular, we use two approaches here: (a) Approach 1, suitable for a moderate number of points, with comparable efficiency to polynomial quadrature rules (leading to about 6 or 7 digits in accuracy), and (b) Approach 2, suitable for a large number of points but less efficient (leading to machine accuracy).  Symmetric rules that can efficiently handle singularities are desirable because their mapping to the integration domain is straightforward and points are not heavily concentrated near some vertices.  Asymmetric rules, on the other hand, generally employed to integrate singularities, require the determination of vertex mapping, and points may be concentrated nonuniformly at the vertices.

This paper is organized as follows.  In Section~\ref{sec:singularity}, we characterize the singularities in the test-element integrand.  In Section~\ref{sec:quadrature}, we use the characterizations from Section~\ref{sec:singularity} to construct appropriate geometrically symmetric quadrature rules.  In Section~\ref{sec:results}, we demonstrate the effectiveness of these rules and compare them to a standard polynomial scheme and the DMRW rules.  In Section~\ref{sec:conclusions}, we provide concluding remarks.

%% file: singularity.tex
\section{Logarithmic Singularities in the MoM Test Integrand for the EFIE}
\label{sec:singularity}

Singularities will appear in the source potential when it becomes the integrand of a test integral.  Under the $e^{j\omega t}$ time-harmonic convention, the singular integrals for the EFIE that occur when using the MoM take the forms
\begin{gather}
I_s = \int_{A_\test}\nabla\cdot\boldsymbol{\Lambda}_\test^j(\mathbf{x}_\test) \int_{A_\src} \frac{e^{-jkR(\mathbf{x}_\src,\mathbf{x}_\test)}}{R(\mathbf{x}_\src,\mathbf{x}_\test)}\nabla\cdot\boldsymbol{\Lambda}_\src^i(\mathbf{x}_\src) dA_\src dA_\test
\label{eq:esp}
\end{gather}
and
\begin{align}
I_v = \int_{A_\test}\boldsymbol{\Lambda}_\test^j(\mathbf{x}_\test)\cdot 
\int_{A_\src} \frac{e^{-jkR(\mathbf{x}_\src,\mathbf{x}_\test)}}{R(\mathbf{x}_\src,\mathbf{x}_\test)}\boldsymbol{\Lambda}_\src^i(\mathbf{x}_\src) dA_\src dA_\test,
\label{eq:evp}
\end{align}
where $I_s$ in~\eqref{eq:esp} appears in the scalar potential in the EFIE and $I_v$ in~\eqref{eq:evp} appears in the vector potential in the EFIE.  $\mathbf{x}_\src$ and $\mathbf{x}_\test$ are the source and test points, respectively; $A_\src$ and $A_\test$ are the source and test elements surfaces, respectively; $R(\mathbf{x}_\src,\mathbf{x}_\test) = \|\mathbf{x}_\src-\mathbf{x}_\test\|_2$, $\boldsymbol{\Lambda}_\test^j$ is the test basis function associated with edge $j$; and $\boldsymbol{\Lambda}_\src^i$ is the source basis function associated with edge $i$.  In~\eqref{eq:esp} and~\eqref{eq:evp}, $k = k_0\sqrt{\epsilon_r\mu_r}$, where $k_0=2\pi/\lambda$ is the free-space wavenumber, $\lambda$ is the wavelength, and $\epsilon_r$ and $\mu_r$ are the relative permittivity and permeability of the medium, respectively.

When $\boldsymbol{\Lambda}_\test^j$ and $\boldsymbol{\Lambda}_\src^i$ are linear, as in the Rao--Wilton--Glisson (RWG) basis functions~\cite{rao_1982}, $\nabla\cdot\boldsymbol{\Lambda}_\test^j(\mathbf{x}_\test)$ and $\nabla\cdot\boldsymbol{\Lambda}_\src^i(\mathbf{x}_\src)$ are constants, such that~\eqref{eq:esp} becomes
\begin{align}
I_s = C_1\int_{A_\test}\int_{A_\src} \frac{e^{-jkR(\mathbf{x}_\src,\mathbf{x}_\test)}}{R(\mathbf{x}_\src,\mathbf{x}_\test)} dA_\src dA_\test.
\label{eq:esp2}
\end{align}
Upon performing a Taylor-series expansion of the exponential factor about $R$, the test integrand in~\eqref{eq:esp2} can be expressed as
\begin{align}
f(\mathbf{x}_\test) ={}& \sum_{p=0}^\infty \frac{(-jk)^p}{p!} \int_{A_\src} R(\mathbf{x}_\src,\mathbf{x}_\test)^{p-1} dA_\src. 
\label{eq:taylor_series}
\end{align}
Even $p$ terms in~\eqref{eq:taylor_series}, which raise $R$ to odd powers, yield terms with unbounded derivatives near the boundaries of $A_\src$.  Odd $p$ terms in~\eqref{eq:taylor_series} yield even powers of $R$, which remain smooth and integrable.  

As in~\cite{rao_1982}, when $\boldsymbol{\Lambda}_\test^j$ and $\boldsymbol{\Lambda}_\src^i$ are unnormalized RWG basis functions of the form $\boldsymbol{\Lambda}_\test^j(\mathbf{x}_\test)=\mathbf{x}_\test-\mathbf{x}_j$ and $\boldsymbol{\Lambda}_\src^i(\mathbf{x}_\src)=\mathbf{x}_\src-\mathbf{x}_i$, where $\mathbf{x}_j$ is the vertex of the test element opposite edge $j$ and $\mathbf{x}_i$ is the vertex of the source element opposite edge $i$, $\boldsymbol{\Lambda}_\test^j\cdot\boldsymbol{\Lambda}_\src^i$ in~\eqref{eq:evp} becomes
\begin{align}
\boldsymbol{\Lambda}_\test^j\cdot\boldsymbol{\Lambda}_\src^i
=
(\mathbf{x}_\test-\mathbf{x}_j)\cdot(\mathbf{x}_\src-\mathbf{x}_i)
=
\left(\tilde{\mathbf{x}}+\frac{\mathbf{x}_\test-\mathbf{x}_\src}{2}-\mathbf{x}_j\right)\cdot\left(\tilde{\mathbf{x}}-\frac{\mathbf{x}_\test-\mathbf{x}_\src}{2}-\mathbf{x}_i\right)
=
D_0+D_1 R +D_2 R^2,
\label{eq:quadratic}
\end{align}
where
\begin{align*}
D_0(\mathbf{x}_\src,\mathbf{x}_\test) ={}&\|\tilde{\mathbf{x}}\|_2^2-(\mathbf{x}_i+\mathbf{x}_j)\cdot\tilde{\mathbf{x}}+\mathbf{x}_i\cdot\mathbf{x}_j, \\[.5em]
D_1(\mathbf{x}_\src,\mathbf{x}_\test) ={}&\frac{\|\mathbf{x}_j-\mathbf{x}_i\|_2}{2}\cos \phi(\mathbf{x}_\src,\mathbf{x}_\test), \\[.5em]
D_2 ={}& -1/4,
\end{align*}
$\tilde{\mathbf{x}}(\mathbf{x}_\src,\mathbf{x}_\test)=(\mathbf{x}_\src+\mathbf{x}_\test)/2$, and $\phi(\mathbf{x}_\src,\mathbf{x}_\test)$ is the angle between $(\mathbf{x}_\test-\mathbf{x}_\src)$ and $(\mathbf{x}_j-\mathbf{x}_i)$.
Using~\eqref{eq:quadratic}, \eqref{eq:evp} becomes
\begin{align}
I_v ={}& \int_{A_\test} \int_{A_\src} D_0(\mathbf{x}_\src,\mathbf{x}_\test)\frac{e^{-jkR(\mathbf{x}_\src,\mathbf{x}_\test)}}{R(\mathbf{x}_\src,\mathbf{x}_\test)} dA_\src dA_\test 
{}+{}
\int_{A_\test} \int_{A_\src} D_1(\mathbf{x}_\src,\mathbf{x}_\test)e^{-jkR(\mathbf{x}_\src,\mathbf{x}_\test)} dA_\src dA_\test \nonumber \\
&{}+{}
D_2\int_{A_\test} \int_{A_\src} e^{-jkR(\mathbf{x}_\src,\mathbf{x}_\test)}R(\mathbf{x}_\src,\mathbf{x}_\test) dA_\src dA_\test.
\label{eq:evp2}
\end{align}
Performing a Taylor series expansion of the exponential factor in~\eqref{eq:evp2} leads to integer powers of $R$.  
Once more, odd powers of $R$ yield singularities, whereas even powers remain smooth and integrable.


We describe the singularities 
arising from the odd powers of $R$ in
\begin{gather}
\int_{A_\src}R(\mathbf{x}_\src,\mathbf{x}_\test)^q dA_\src, \quad \text{for } q=-1,\,0,\,1,\hdots
\label{eq:singular}
\end{gather}
for two cases: \rereading{(1) when $A_\src$ and $A_\test$ are coplanar, and (2) when $A_\src$ and $A_\test$ are perpendicular and share an edge.}
Note that, although~\eqref{eq:singular} is useful to understand and discuss singularities, integrations will not be carried out using~\eqref{eq:singular}; rather, the more general integrals in~\eqref{eq:esp} and~\eqref{eq:evp} will be used in later sections. 
Though the primary focus of this paper is on triangular elements, for simplicity, we use here rectangular domains to describe the singularities so that the equations are more tractable.  However, we have also investigated these cases for the triangular elements and, though the expressions are more complicated, they retain the same singularities as the rectangular elements shown below.

\subsection{Coplanar Domains} 
\label{sec:coplanar}

\begin{figure}[htbp]
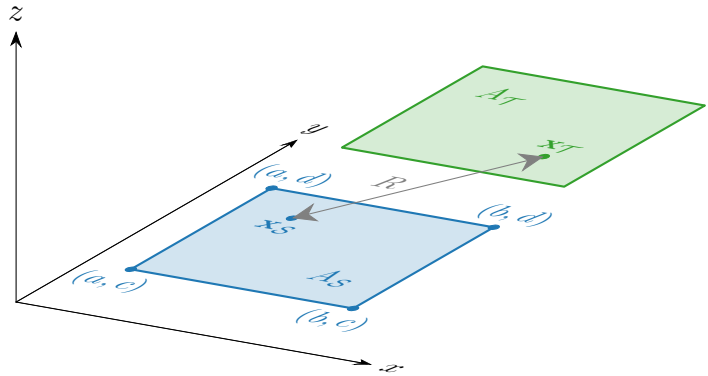

\centering
\usebox{\coplanarrectangles}%
\caption{Coplanar domains: $A_\src$ and $A_\test$.}
\label{fig:coplanar_rectangles}
\end{figure}

In this subsection, we analyze the kind of singularities exhibited for coplanar domains. 
We demonstrate this behavior by letting $A_\src$ be the rectangle $\xsrc\in[a,b]\times\ysrc\in[c,d]$ and $A_\test$ be coplanar with $A_\src$.  \rereading{These domains are shown in Fig.~\ref{fig:coplanar_rectangles}.}

\input{fig_coplanar_rectangles.tex}

For $q=-1$ in~\eqref{eq:singular}, $\int_{A_\src} R(\mathbf{x}_\src,\mathbf{x}_\test)^{-1} dA_\src$ can be computed with the aid of a computer algebra system, such as~\cite{mathematica}, to show that
\begin{gather}
\int_{c}^{d}\int_{a}^{b} \frac{1}{\sqrt{(\xtest-\xsrc)^2+(\ytest-\ysrc)^2}} d\xsrc d\ysrc =
\sum_{i=1}^4\left\{
\alpha_i\ln\left[\beta_i+\sqrt{\alpha_i^2+\beta_i^2}\right] 
{}-{} 
\alpha_i\ln\left[\gamma_i+\sqrt{\alpha_i^2+\gamma_i^2}\right]\right\} ,
\label{eq:p_0}
\end{gather}
where 
\begin{align*}
\alpha &{}= \{\ytest-c,\,\ytest-d,\,\xtest-a,\,\xtest-b\},
\\
\beta &{}= \{\xtest-a,\,\xtest-b,\,\ytest-c,\,\ytest-d\},
\\
\gamma &{}= \{\xtest-b,\,\xtest-a,\,\ytest-d,\,\ytest-c\}.
\end{align*}

In~\eqref{eq:p_0}, each pair of terms in the summation yields singularities in derivatives with respect to the test coordinates along $\alpha_i=0$.
%
%
Approaching the edges, where $\beta_i=0$ or $\gamma_i=0$, the two terms in~\eqref{eq:p_0} become
\begin{align}
\lim_{\beta_i \to 0} \alpha_i\ln\left[ \beta_i+\sqrt{\alpha_i^2+ \beta_i^2}\right] &{}= \alpha_i\ln|\alpha_i|, \label{eq:p_0_2d_beta}\\
\lim_{\gamma_i\to 0} \alpha_i\ln\left[\gamma_i+\sqrt{\alpha_i^2+\gamma_i^2}\right] &{}= \alpha_i\ln|\alpha_i|. \label{eq:p_0_2d_gamma}
\end{align}
At the vertices, where $\alpha_i=0$, $\alpha_i\ln|\alpha_i|$ is singular.
%
\rereading{Additionally, on an edge of the rectangle not at a vertex, $\alpha_i=0$ and $\beta_i$ and $\gamma_i$ are nonzero with opposite signs.  Taylor series expansions of the arguments of the logarithms are
\begin{alignat}{11}
   \beta_i&{}+\sqrt{\alpha_i^2+ \beta_i^2} &{}=  \beta_i &&{}+| \beta_i| &&{}+ \frac{\alpha_i^2}{2| \beta_i|}&&{}+\mathcal{O}(\alpha_i^3), \label{eq:log_arg_beta} \\
  \gamma_i&{}+\sqrt{\alpha_i^2+\gamma_i^2} &{}= \gamma_i &&{}+|\gamma_i| &&{}+ \frac{\alpha_i^2}{2|\gamma_i|}&&{}+\mathcal{O}(\alpha_i^3). \label{eq:log_arg_gamma}
\end{alignat}
When $\beta_i<0$, \eqref{eq:log_arg_beta} is approximately $\frac{\alpha_i^2}{2| \beta_i|}$.  When $\gamma_i<0$, \eqref{eq:log_arg_gamma} is approximately $\frac{\alpha_i^2}{2| \gamma_i|}$.  Either of these conditions yields a term containing $\alpha_i\ln|\alpha_i|$, indicating the edge has a singularity as well.}
%
\rereading{Approaching an edge of the rectangle, where $\alpha_i=0$,} a series expansion of~\eqref{eq:p_0} as $\alpha_i\to0$ yields the following terms:
\begin{align}
1,\,\alpha_i,\,\alpha_i\ln |\alpha_i|,\,\alpha_i^2,\,\alpha_i^3,\,\alpha_i^4,\,\alpha_i^5,\hdots.
\label{eq:p_0_1d_series}
\end{align}
The series in~\eqref{eq:p_0_1d_series} consists of monomial terms, as expected from a Taylor series expansion, as well as the aforementioned $\alpha_i\ln|\alpha_i|$ term.

For $q=1$ in~\eqref{eq:singular}, $\int_{A_\src} R(\mathbf{x}_\src,\mathbf{x}_\test) dA_\src$ yields additional terms, however, similar analysis yields similar observations.  The terms
\begin{gather*}
\alpha_i^3\ln\left[\beta_i+\sqrt{\alpha_i^2+\beta_i^2}\right], \qquad \alpha_i^3\ln\left[\gamma_i+\sqrt{\alpha_i^2+\gamma_i^2}\right]
\end{gather*}
yield singularities of the form $\alpha_i^3\ln|\alpha_i|$ (similar to \eqref{eq:p_0_2d_beta} and \eqref{eq:p_0_2d_gamma}) along the edges and vertices and, along the edge, a series expansion of $\int_{A_\src} R(\mathbf{x}_\src,\mathbf{x}_\test) dA_\src$ as $\alpha_i\to0$ yields the following terms:
\begin{align}
1,\,\alpha_i,\,\alpha_i^2,\,\alpha_i^3,\,\alpha_i^3\ln |\alpha_i|,\,\alpha_i^4,\,\alpha_i^5,\hdots.
\label{eq:p_2_1d_series}
\end{align}
The series in~\eqref{eq:p_2_1d_series} consists of monomial terms, as expected from a Taylor series expansion, as well as the $\alpha_i^3\ln|\alpha_i|$ term.

The trend continues for odd powers of $R$, with one-dimensional characterizations $\alpha_i^{q+2}\ln|\alpha_i|$ and 2D characterizations
\begin{gather}
\alpha_i^{q+2}\ln\left[\beta_i+\sqrt{\alpha_i^2+\beta_i^2}\right], \qquad
\alpha_i^{q+2}\ln\left[\gamma_i+\sqrt{\alpha_i^2+\gamma_i^2}\right].
\label{eq:2d_trend}
\end{gather}
For even powers of $R$, the test integrand consists of a linear combination of monomials $\xtest^s \ytest^t$, where $0\le s \le q$, $0\le t \le q$, and $0\le s+t \le q$.

These derivations indicate that, when $A_\test=A_\src$, the entire boundary of $A_\test$ has singularities.  When $A_\test$ and $A_\src$ are co-planar and share an edge, the shared edge and its vertices have singularities.  The numerical procedure here developed is applicable to both self and touching elements.

\subsection{Perpendicular Domains} 
\label{sec:perpendicular}
\begin{figure}[htbp]
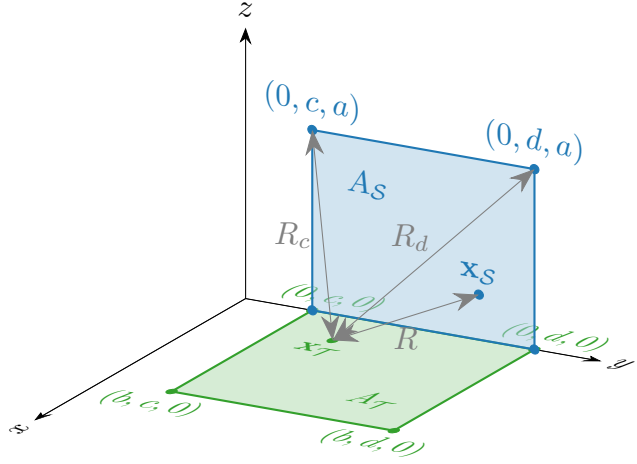

\centering
\usebox{\perpendicularrectangles}%
\vspace{-0.5em}
\caption{Perpendicular domains: $A_\src$ and $A_\test$.}
\label{fig:perpendicular_rectangles}
\vspace{-1em}
\end{figure}

In this subsection, we analyze the kind of singularities exhibited in perpendicular domains. 
We demonstrate this behavior by letting $A_\src$ be the rectangle $\ysrc\in[c,d]\times\zsrc\in[0,a]$ on the $x=0$ plane and $A_\test$ be the rectangle $\xtest\in[0,b]\times\ytest\in[c,d]$ on the $z=0$ plane (note the shared edge is along the $y$-axis, where $\xtest=0$ and $\zsrc=0$).  \rereading{These domains are shown in Fig.~\ref{fig:perpendicular_rectangles}.}

\input{fig_perpendicular_rectangles.tex}

For $q=-1$ in~\eqref{eq:singular}, $\int_{A_\src} R(\mathbf{x}_\src,\mathbf{x}_\test)^{-1} dA_\src$ can be computed with the aid of a computer algebra system, such as~\cite{mathematica}, to show that
%
\rereading{%
\begin{align}
\int_{0}^{a}\int_{c}^{d} \frac{1}{\sqrt{\xtest^2+(\ytest-\ysrc)^2+\zsrc^2}} d\ysrc d\zsrc =&{} 
{}-\xtest\arctan\frac{a y_c}{\xtest R_c} 
{}+\xtest\arctan\frac{a y_d}{\xtest R_d}  \nonumber\\[.5em]
&{}-\frac{y_c}{2}\ln\left[\xtest^2+y_c^2\right] 
{}+\frac{y_d}{2}\ln\left[\xtest^2+y_d^2\right]  \nonumber\\[.5em]
&{}+y_c\ln\left[a+R_c\right] 
{}-y_d\ln\left[a+R_d\right]  \nonumber\\[.5em]
&{}+a\ln\left[y_c+R_c\right] 
{}-a\ln\left[y_d+R_d\right],
\label{eq:p_0_perp}
\end{align}}%
%
\rereading{where $R_c^2 = \xtest^2 + (\ytest-c)^2 + a^2$, $R_d^2 = \xtest^2 + (\ytest-d)^2 + a^2$, $y_c = \ytest-c$, and $y_d = \ytest-d$.}

In~\eqref{eq:p_0_perp}, there are unbounded derivatives at the shared vertices $(0,c,0)$ and $(0,d,0)$ that respectively arise from \rereading{$\frac{y_c}{2}\ln\left[\xtest^2+y_c^2\right]$ and $\frac{y_d}{2}\ln\left[\xtest^2+y_d^2\right]$.  As $\xtest\to 0$, these become $\alpha_i\ln|\alpha_i|$, where $\alpha=\{y_c,\,y_d\}$.}  A series expansion of $\int_{A_\src} R(\mathbf{x}_\src,\mathbf{x}_\test)^{-1} dA_\src$ from the interior of $A_\test$ to a shared vertex yields the following terms:
\begin{align*}
1,\,\alpha_i,\,\alpha_i\ln |\alpha_i|,\,\alpha_i^2,\,\alpha_i^3,\,\alpha_i^4,\,\alpha_i^5,\hdots.
\end{align*}

As with the coplanar case in Section~\ref{sec:coplanar}, for even powers of $R$, the test integrand consists of a linear combination of monomials $\xtest^s \ytest^t$, where $0\le s \le q$, $0\le t \le q$, and $0\le s+t \le q$.

For $q=1$ in~\eqref{eq:singular}, $\int_{A_\src} R(\mathbf{x}_\src,\mathbf{x}_\test) dA_\src$ yields additional terms, however, similar analysis yields similar observations.  The terms
\rereading{%
\begin{align*}
\frac{\left(3\xtest^2+y_c^2\right)y_c}{12}\ln[\xtest^2+y_c^2], \qquad
\frac{\left(3\xtest^2+y_d^2\right)y_d}{12}\ln[\xtest^2+y_d^2]
\end{align*}}%
yield singularities of the form $\alpha_i^3\ln|\alpha_i|$ as $\xtest\to 0$.

A series expansion of $\int_{A_\src} R(\mathbf{x}_\src,\mathbf{x}_\test) dA_\src$ from the interior of $A_\test$ to a shared vertex yields the following terms:
\begin{align*}
1,\,\alpha_i,\,\alpha_i^2,\,\alpha_i^3,\,\alpha_i^3\ln |\alpha_i|,\,\alpha_i^4,\,\alpha_i^5,\hdots.
\end{align*}

The trend continues for the odd powers of $R$, with one-dimensional characterizations $\alpha_i^{q+2}\ln|\alpha_i|$.
These derivations indicate that, when $A_\src$ and $A_\test$ are perpendicular with a shared edge, the shared vertices have singularities.  

\rereading{It should also be noted that the arctangent terms in \eqref{eq:p_0_perp} yield unbounded derivatives at the shared vertices as well.  These singularities behave differently from the logarithmic singularities, in that they do not introduce singular terms in the series expansions.  Nonetheless, our approaches in the next section for the latter mitigate the severity of the former.}

%% file: fig_coplanar_rectangles.tex
\newsavebox{\coplanarrectangles}%
\sbox{\coplanarrectangles}{%
\newcommand*\projA{-20}
\newcommand*\projB{ 35}
%
\definecolor{darkred} {RGB}{227,26,28}%
\definecolor{darkorange} {RGB}{255,127,0}%
\definecolor{lightred} {RGB}{243.8,163.4,164.2}%
\definecolor{darkblue}  {RGB}{ 31,120,180}%
\definecolor{lightblue}  {RGB}{165.4,201,225}
\definecolor{darkgreen} {RGB}{ 51,160, 44}%
\definecolor{lightgreen}{RGB}{173.4,217,170.6}
\def\xangle{100}
\def\yangle{60}
\begin{tikzpicture}[scale=0.6,
x={({ sin(\xangle)*1cm},{ cos(\xangle)*1cm})}, 
y={({ .9*sin(\yangle)*1cm},{ .9*cos(\yangle)*1cm})}, 
z={(0cm,1cm)}
    ]   
   
   \def\Deltaz{0};    
   \def\Deltay{2};    
   \def\Deltaysrc{2}; 
   \def\yb{0};        
   \def\yt{4};        
   \def\xl{1};        
   \def\xr{6};        
   \def\meas{1};      
   \def\mw{.25};      
   \def\thetaval{0}; 
   \def\arcrad{1};    
   \def\ep{.1}        
   \def\ymax{8}   
   
   \useasboundingbox (-1cm,-2cm) rectangle (16cm,7cm);
   
   \begin{scope}[canvas is xy plane at z=0,transform shape]
   \draw[line cap=round] (0,0) -- (8-\ep, 0);
   \draw[line cap=round] (0,0) -- (0,\ymax-\ep);
   \pgflowlevelsynccm
   \draw[-{Stealth[scale=2.0]},line cap=round] (0,0) to (8, 0) node[right,scale=2,dash pattern=on 0cm off 100cm] {$x$};
   \draw[-{Stealth[scale=2.0]},line cap=round] (0,0) to (0,\ymax) node[above,scale=2,dash pattern=on 0cm off 100cm] {$y$};
   \end{scope}

   \def\xsx{\xl*.75+\xr*.25};
   \def\xsy{\yt*.75+\yb*.25+\Deltaysrc};
   \begin{scope}[canvas is xy plane at z=0]
      \draw[line join=round,thick,darkblue,fill=lightblue,fill opacity=.5,draw opacity=1] 
       (\xr,\yb+\Deltaysrc) -- 
       (\xr,\yt+\Deltaysrc) -- 
       (\xl,\yt+\Deltaysrc) -- 
       (\xl,\yb+\Deltaysrc) -- cycle;     
      \node[transform shape,scale=2,text=darkblue] at (\xl*.25+\xr*.75,\yt*.25+\yb*.75+\Deltaysrc) {$A_\src$};

      
      \coordinate (xs) at (\xsx,\xsy);
      \draw[draw=darkblue,fill=darkblue] (xs) circle (.1);
      \node[transform shape,scale=2,text=darkblue,anchor=north] at (xs) {$\mathbf{x}_\src$};
      
      \node[transform shape,scale=2,text=darkblue,anchor=north] at (\xl,\yb+\Deltaysrc) {$(a,c)$};
      \node[transform shape,scale=2,text=darkblue,anchor=north] at (\xr,\yb+\Deltaysrc) {$(b,c)$};
      \node[transform shape,scale=2,text=darkblue,anchor=south] at (\xl,\yt+\Deltaysrc) {$(a,d)$};
      \node[transform shape,scale=2,text=darkblue,anchor=south] at (\xr,\yt+\Deltaysrc) {$(b,d)$};
      
      \draw[draw=darkblue,fill=darkblue] (\xl,\yb+\Deltaysrc) circle (.1);
      \draw[draw=darkblue,fill=darkblue] (\xr,\yb+\Deltaysrc) circle (.1);
      \draw[draw=darkblue,fill=darkblue] (\xl,\yt+\Deltaysrc) circle (.1);
      \draw[draw=darkblue,fill=darkblue] (\xr,\yt+\Deltaysrc) circle (.1);
   \end{scope} 

   \def\xtx{\xl*.25+\xr*.75};
   \def\xty{\yt*.25+\yb*.75};
   \begin{scope}[canvas is plane={%
   O(0,{\yt+\Deltay+\Deltaysrc               },{\Deltaz               })%
   x(1,{\yt+\Deltay+\Deltaysrc               },{\Deltaz               })%
   y(0,{\yt+\Deltay+\Deltaysrc+cos(\thetaval)},{\Deltaz+sin(\thetaval)})}]
      
      \draw[line join=round,thick,draw=darkgreen,fill=lightgreen,fill opacity=.5,draw opacity=1] (\xr,\yb) -- (\xr,\yt) -- (\xl,\yt) -- (\xl,\yb) -- cycle;
      
      \node[transform shape,scale=2,text=darkgreen] at (\xl*.75+\xr*.25,\yt*.75+\yb*.25) {$A_\test$};
      
      
      
      \coordinate (xt) at (\xtx,\xty);
      \draw[draw=darkgreen,fill=darkgreen] (xt) circle (.1);
      \node[transform shape,scale=2,text=darkgreen,anchor=south] at (xt) {$\mathbf{x}_\test$};

      \node[transform shape,anchor=east,scale=2,inner sep=5pt,gray] at (\xl*.5+\xr*.5,\yb-\Deltay*.5) {$R$};
   \end{scope}
   
   \def\Rax{(\xsx)};
   \def\Rbx{(\xtx)};
   \def\Ray{(\xsy)};
   \def\Rby{(\yt+\Deltay+\Deltaysrc-\yb+(\xty)*cos(\thetaval))};
   \def\Raz{(0)};
   \def\Rbz{(\Deltaz+(\xty)*sin(\thetaval))};
   
   \def\R{sqrt( (\Rbx-\Rax)^2 + (\Rby-\Ray)^2 + (\Rbz-\Raz)^2};
   
   \draw[line cap=round,gray] 
      (
       {\Rax+\ep*(\Rbx-\Rax)/\R}, 
       {\Ray+\ep*(\Rby-\Ray)/\R},
       {\Raz+\ep*(\Rbz-\Raz)/\R}
      ) 
      -- 
      (
       {\Rbx+\ep*(\Rax-\Rbx)/\R}, 
       {\Rby+\ep*(\Ray-\Rby)/\R},  
       {\Rbz+\ep*(\Raz-\Rbz)/\R}
      ); 
   
   \draw[{Stealth[scale=2.0,fill=gray,gray]}-{Stealth[scale=2.0,fill=gray,gray]},line cap=round,dash pattern=on 0cm off 100cm] 
   ({\Rax},{\Ray},{\Raz}) -- ({\Rbx},{\Rby},{\Rbz}) ;
   
   \begin{scope}[canvas is xz plane at y=0]
   \draw[line cap=round] (0,0) -- (0,6-\ep);
   \pgflowlevelsynccm
   \draw[-{Stealth[scale=2.0]},line cap=round,dash pattern=on 0cm off 100cm] (0,0) to (0,6) node[above,scale=2] {$z$};
   \end{scope}

\end{tikzpicture}

}

%
%
%
%
%

%% file: fig_perpendicular_rectangles.tex
\newsavebox{\perpendicularrectangles}%
\sbox{\perpendicularrectangles}{%
\newcommand*\projA{-20}
\newcommand*\projB{ 35}
%
\definecolor{darkred} {RGB}{227,26,28}%
\definecolor{darkorange} {RGB}{255,127,0}%
\definecolor{lightred} {RGB}{243.8,163.4,164.2}%
\definecolor{darkblue}  {RGB}{ 31,120,180}%
\definecolor{lightblue}  {RGB}{165.4,201,225}
\definecolor{darkgreen} {RGB}{ 51,160, 44}%
\definecolor{lightgreen}{RGB}{173.4,217,170.6}
\def\xangle{100}
\def\yangle{60}
\begin{tikzpicture}[scale=0.6,
x={({ sin(\xangle)*1cm},{ cos(\xangle)*1cm})}, 
y={({.9*sin(\yangle)*1cm},{.9*cos(\yangle)*1cm})}, 
z={(0cm,1cm)}
    ]   
   
   \def\Deltaz{0};    
   \def\Deltay{0};    
   \def\Deltaysrc{-4}; 
   \def\yb{0};        
   \def\yt{4};        
   \def\xl{1.5};        
   \def\xr{6.5};        
   \def\meas{1};      
   \def\mw{.25};      
   \def\thetaval{90}; 
   \def\arcrad{1};    
   \def\ep{.1}        
   \def\ymax{6}   
   
   \useasboundingbox (-6cm,-4cm) rectangle (9.5cm,7cm);
   
   \begin{scope}[canvas is xy plane at z=0,transform shape]
   \draw[line cap=round] (0,0) -- (8-\ep, 0);
   \draw[line cap=round] (0,0) -- (0,-\ymax+\ep);
   \pgflowlevelsynccm
   \draw[-{Stealth[scale=2.0]},line cap=round] (0,0) to (8, 0) node[right,scale=2,dash pattern=on 0cm off 100cm] {$y$};
   \draw[-{Stealth[scale=2.0]},line cap=round] (0,0) to (0,-\ymax) node[right,scale=2,dash pattern=on 0cm off 100cm,rotate=270] {$x$};
   \end{scope}

   \def\xsx{\xl*.75+\xr*.25};
   \def\xsy{\yt*.75+\yb*.25+\Deltaysrc};
   \begin{scope}[canvas is xy plane at z=0]
      \draw[line join=round,thick,darkgreen,fill=lightgreen,fill opacity=.5,draw opacity=1] 
       (\xr,\yb+\Deltaysrc) -- 
       (\xr,\yt+\Deltaysrc) -- 
       (\xl,\yt+\Deltaysrc) -- 
       (\xl,\yb+\Deltaysrc) -- cycle;     
      \node[transform shape,scale=2,text=darkgreen] at (\xl*.25+\xr*.75,\yt*.25+\yb*.75+\Deltaysrc) {$A_\test$};

      
      \coordinate (xs) at (\xsx,\xsy);
      \draw[draw=darkgreen,fill=darkgreen] (xs) circle (.1);
      \node[transform shape,scale=2,text=darkgreen,anchor=north] at (xs) {$\mathbf{x}_\test$};
      
      \node[transform shape,scale=2,text=darkgreen,anchor=north] at (\xl,\yb+\Deltaysrc) {$(b,c,0)$};
      \node[transform shape,scale=2,text=darkgreen,anchor=north] at (\xr,\yb+\Deltaysrc) {$(b,d,0)$};
      \node[transform shape,scale=2,text=darkgreen,anchor=south] at (\xl,\yt+\Deltaysrc) {$(0,c,0)$};
      \node[transform shape,scale=2,text=darkgreen,anchor=south] at (\xr,\yt+\Deltaysrc) {$(0,d,0)$};
      
      \draw[draw=darkgreen,fill=darkgreen] (\xl,\yb+\Deltaysrc) circle (.1);
      \draw[draw=darkgreen,fill=darkgreen] (\xr,\yb+\Deltaysrc) circle (.1);
      \draw[draw=darkgreen,fill=darkgreen] (\xl,\yt+\Deltaysrc) circle (.1);
      \draw[draw=darkgreen,fill=darkgreen] (\xr,\yt+\Deltaysrc) circle (.1);
   \end{scope} 

   \def\xtx{\xl*.25+\xr*.75};
   \def\xty{\yt*.25+\yb*.75};
   \begin{scope}[canvas is plane={%
   O(0,{\yt+\Deltay+\Deltaysrc               },{\Deltaz               })%
   x(1,{\yt+\Deltay+\Deltaysrc               },{\Deltaz               })%
   y(0,{\yt+\Deltay+\Deltaysrc+cos(\thetaval)},{\Deltaz+sin(\thetaval)})}]
      
      \draw[line join=round,thick,draw=darkblue,fill=lightblue,fill opacity=.5,draw opacity=1] (\xr,\yb) -- (\xr,\yt) -- (\xl,\yt) -- (\xl,\yb) -- cycle;
      
      \node[transform shape,scale=2,text=darkblue] at (\xl*.75+\xr*.25,\yt*.75+\yb*.25) {$A_\src$};
      
      
      
      \coordinate (xt) at (\xtx,\xty);
      \draw[draw=darkblue,fill=darkblue] (xt) circle (.1);
      \node[transform shape,scale=2,text=darkblue,anchor=south] at (xt) {$\mathbf{x}_\src$};
      
      \node[transform shape,scale=2,text=darkblue,anchor=south] at (\xl,\yt) {$(0,c,a)$};
      \node[transform shape,scale=2,text=darkblue,anchor=south] at (\xr,\yt) {$(0,d,a)$};
      
      \draw[draw=darkblue,fill=darkblue] (\xl,\yb) circle (.1);
      \draw[draw=darkblue,fill=darkblue] (\xr,\yb) circle (.1);
      \draw[draw=darkblue,fill=darkblue] (\xl,\yt) circle (.1);
      \draw[draw=darkblue,fill=darkblue] (\xr,\yt) circle (.1);
      
   \end{scope}
   
   \def\Rax{(\xsx)};
   \def\Rbx{(\xtx)};
   \def\Ray{(\xsy)};
   \def\Rby{(\yt+\Deltay+\Deltaysrc-\yb+(\xty)*cos(\thetaval))};
   \def\Raz{(0)};
   \def\Rbz{(\Deltaz+(\xty)*sin(\thetaval))};
   
   \def\R{sqrt( (\Rbx-\Rax)^2 + (\Rby-\Ray)^2 + (\Rbz-\Raz)^2};
   
   \draw[line cap=round,gray] 
      (
       {\Rax+\ep*(\Rbx-\Rax)/\R}, 
       {\Ray+\ep*(\Rby-\Ray)/\R},
       {\Raz+\ep*(\Rbz-\Raz)/\R}
      ) 
      -- 
      (
       {\Rbx+\ep*(\Rax-\Rbx)/\R}, 
       {\Rby+\ep*(\Ray-\Rby)/\R},  
       {\Rbz+\ep*(\Raz-\Rbz)/\R}
      ); 
   
   \draw[{Stealth[scale=2.0,fill=gray,gray]}-{Stealth[scale=2.0,fill=gray,gray]},line cap=round,dash pattern=on 0cm off 100cm] 
   ({\Rax},{\Ray},{\Raz}) -- ({\Rbx},{\Rby},{\Rbz}) node[transform shape,midway,anchor=north,scale=2,inner sep=3pt,gray] {$R$};
   
   \def\Rax{(\xsx)};
   \def\Rbx{(\xl)};
   \def\Ray{(\xsy)};
   \def\Rby{(\yt+\Deltay+\Deltaysrc-\yb+(\yt)*cos(\thetaval))};
   \def\Raz{(0)};
   \def\Rbz{(\Deltaz+(\yt)*sin(\thetaval))};
   
   \def\R{sqrt( (\Rbx-\Rax)^2 + (\Rby-\Ray)^2 + (\Rbz-\Raz)^2};
   
   \draw[line cap=round,gray] 
      (
       {\Rax+\ep*(\Rbx-\Rax)/\R},
       {\Ray+\ep*(\Rby-\Ray)/\R},
       {\Raz+\ep*(\Rbz-\Raz)/\R}
      ) 
      -- 
      (
       {\Rbx+\ep*(\Rax-\Rbx)/\R}, 
       {\Rby+\ep*(\Ray-\Rby)/\R},  
       {\Rbz+\ep*(\Raz-\Rbz)/\R}
      ); 
   
   \draw[{Stealth[scale=2.0,fill=gray,gray]}-{Stealth[scale=2.0,fill=gray,gray]},line cap=round,dash pattern=on 0cm off 100cm] 
   ({\Rax},{\Ray},{\Raz}) -- ({\Rbx},{\Rby},{\Rbz})  node[transform shape,midway,left,scale=2,gray] {$R_c$};
   
   \def\Rax{(\xsx)};
   \def\Rbx{(\xr)};
   \def\Ray{(\xsy)};
   \def\Rby{(\yt+\Deltay+\Deltaysrc-\yb+(\yt)*cos(\thetaval))};
   \def\Raz{(0)};
   \def\Rbz{(\Deltaz+(\yt)*sin(\thetaval))};
   
   \def\R{sqrt( (\Rbx-\Rax)^2 + (\Rby-\Ray)^2 + (\Rbz-\Raz)^2};
   
   \draw[line cap=round,gray] 
      (
       {\Rax+\ep*(\Rbx-\Rax)/\R},
       {\Ray+\ep*(\Rby-\Ray)/\R},
       {\Raz+\ep*(\Rbz-\Raz)/\R}
      ) 
      -- 
      (
       {\Rbx+\ep*(\Rax-\Rbx)/\R}, 
       {\Rby+\ep*(\Ray-\Rby)/\R},  
       {\Rbz+\ep*(\Raz-\Rbz)/\R}
      ); 
   
   \draw[{Stealth[scale=2.0,fill=gray,gray]}-{Stealth[scale=2.0,fill=gray,gray]},line cap=round,dash pattern=on 0cm off 100cm] 
   ({\Rax},{\Ray},{\Raz}) -- ({\Rbx},{\Rby},{\Rbz})  node[transform shape,midway,anchor=south east,scale=2,inner sep=1pt,gray] {$R_d$};
   
   \begin{scope}[canvas is xz plane at y=0]
   \draw[line cap=round] (0,0) -- (0,6-\ep);
   \pgflowlevelsynccm
   \draw[-{Stealth[scale=2.0]},line cap=round,dash pattern=on 0cm off 100cm] (0,0) to (0,6) node[above,scale=2] {$z$};
   \end{scope}

\end{tikzpicture}

}

%
%
%
%
%

%% file: quadrature_rules.tex
\section{Geometrically Symmetric Quadrature Rules for Logarithmic Singularities}
\label{sec:quadrature}

Due to the edge and vertex singularities described in Section~\ref{sec:singularity}, quadrature rules for polynomials are not well suited for integrating the test integrand.  Therefore, we construct two types of symmetric quadrature rules for triangles, using Approaches 1 and 2 of~\cite{freno_quad}.  To simplify the notation, we write the function sequences such that $x,y\in[0,1]$.

For Approach 1, which provides sufficient accuracy with the least number of quadrature points, we construct a two-dimensional function sequence that consists of monomials and the two-dimensional characterizations from \eqref{eq:2d_trend} in Section~\ref{sec:coplanar}:
\begin{gather}
1,\,x,\,                                            
x\ln\bigl(y-1+\sqrt{x^2+(y-1)^2}\bigr),\,           
x\ln\bigl(y+\sqrt{x^2+y^2}\bigr),\,                
x^2,\,xy,\,                                       
x^3,\,x^2y,\,                                       
x^3\ln\bigl(y-1+\sqrt{x^2+(y-1)^2}\bigr),\,   \nonumber \\     
x^3\ln\bigl(y+\sqrt{x^2+y^2}\bigr),\,               
x^4,\,x^3y,\,x^2y^2,\,                              
x^5,\,x^4y,\,x^3y^2,\,                         
x^5\ln\bigl(y-1+\sqrt{x^2+(y-1)^2}\bigr),\,        
x^5\ln\bigl(y+\sqrt{x^2+y^2}\bigr),\,           \nonumber \\    
x^6,\,x^5y,\,x^4y^2,\,x^3y^3,\,                    
x^7,\,x^6y,\,x^5y^2,\,x^4y^3,\,                    
x^7\ln\bigl(y-1+\sqrt{x^2+(y-1)^2}\bigr),\,         
x^7\ln\bigl(y+\sqrt{x^2+y^2}\bigr),\,               \nonumber \\
x^8,\,x^7y,\,x^6y^2,\,x^5y^3,\,x^4y^4,\,            
x^9,\,x^8y,\,x^7y^2,\,x^6y^3,\,x^5y^4,\,            
x^9\ln\bigl(y-1+\sqrt{x^2+(y-1)^2}\bigr),\,         
x^9\ln\bigl(y+\sqrt{x^2+y^2}\bigr),\,               \nonumber \\
x^{10},\,x^9y,\,x^8y^2,\,x^7y^3,\,x^6y^4,\,x^5y^5,\hdots.
\label{eq:2d_func_seq}
\end{gather}
The points and weights that integrate the function sequence~\eqref{eq:2d_func_seq} are obtained by solving a multidimensional unconstrained optimization problem, as described in~\cite{freno_quad}.
For each of these functions, we map $x\to \alpha$ and $y\to \beta$, where $\alpha$ and $\beta$ are the barycentric coordinates of a triangle (see \ref{app:a}).  Because the rules are geometrically symmetric, these quadrature rules are able to account for the singularities at each edge and vertex.
In this paper, we consider only the two-dimensional characterizations from Section~\ref{sec:coplanar} because these are more severe than those in Section~\ref{sec:perpendicular}.  The points and weights are listed in \ref{app:a}, together with a pictorial representation of the proposed approach.

For Approach 2, which provides monotonic improvement in accuracy as the number of quadrature points is increased, we construct a one-dimensional function sequence
\begin{align}
1,\,x,\,x\ln x,\,x^2,\,x^3,\,x^3\ln x,\,x^4,\,x^5,\,x^5\ln x,\,x^6,\hdots,
\label{eq:1d_func_seq}
\end{align}
which, mapping $x\to \alpha$, follows the trend of \eqref{eq:p_0_1d_series} and \eqref{eq:p_2_1d_series} and applies to the singularities in both Sections~\ref{sec:coplanar} and~\ref{sec:perpendicular}.

The one-dimensional points and weights are listed in \ref{app:b}, for $x\in[0,\,1]$, together with a pictorial representation of the proposed approach.  

Letting $x'=1-x$, Approach 2 can be directly applied to quadrilateral elements by taking the outer product of the one-dimensional rules that exactly integrate
\begin{gather}
1,\,x,\,x\ln x,\,x'\ln x',\,x^2,\,x^3,\,x^3\ln x,\,x'^3\ln x',\,  x^4,\,x^5,\,x^5\ln x,\,x'^5\ln x',\,x^6,\hdots.
\label{eq:1d_func_seq_quad}
\end{gather}
The terms added to \eqref{eq:1d_func_seq} to construct \eqref{eq:1d_func_seq_quad} account for the singularities on the opposite edges.

Although, in this context, we use quadrature rules for logarithmic functions per the singularities observed in Section~\ref{sec:singularity}, Approaches 1 and 2 are well suited to work with the logarithmic singularity in~\eqref{eq:2d_func_seq} and~\eqref{eq:1d_func_seq} replaced by other integrable singular functions.

%% file: results.tex
\section{Numerical Experiments for Singular, Near-Singular, and Far interactions}
\label{sec:results}

\begin{figure}
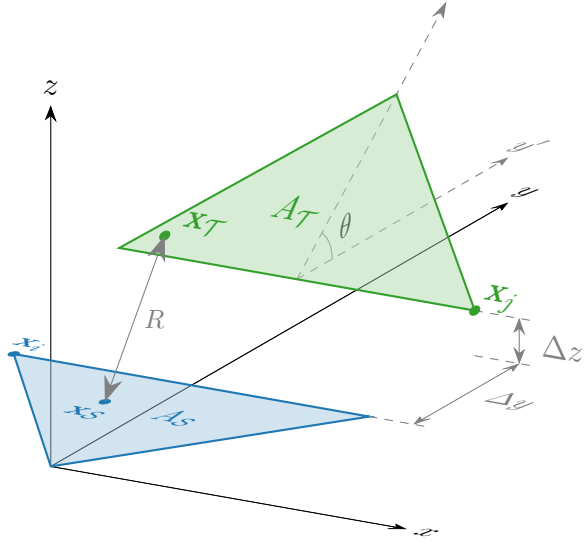

\centering
\usebox{\twotriangles}%
\caption{Relative positions of $A_\src$ and $A_\test$.  The $y'$-axis is on the $x=0$ plane, parallel to the $y$-axis, offset by $\Delta z$.}
\label{fig:triangles}
\end{figure}

To assess the effectiveness of the quadrature rules described in Section~\ref{sec:quadrature}, we consider multiple configurations for $A_\src$ and $A_\test$.  For $A_\src$, we consider the triangular element with vertices $(0\,\mathrm{m},0\,\mathrm{m})$, $(1/20\,\mathrm{m},1/20\,\mathrm{m})$, and $(-1/20\,\mathrm{m},1/20\,\mathrm{m})$, and we use a triangular element with the same shape for $A_\test$. We parameterize these configurations by defining an angle $\theta$ between the planes of $A_\src$ and $A_\test$.  Additionally, we consider displacements $\Delta y$ and $\Delta z$.  These \rereading{elements and} parameters are depicted in Fig.~\ref{fig:triangles}, and are listed in Table~\ref{tab:parameters} for the cases considered in this paper, with $\delta_y=(6\sqrt{2}-1)/60$ m and $\delta_z=1/2000$ m.  
\rereading{Note that we obtain the analog to the coplanar configuration of Section~\ref{sec:coplanar} when $\theta=0^\circ$ or $\theta=180^\circ$, and $\Delta z=0$.  We obtain the analog to the shared-edge perpendicular configuration of Section~\ref{sec:perpendicular} when $\theta=\pm 90^\circ$ and $\Delta y=\Delta z=0$.}

\input{fig_two_triangles.tex}

We consider the integrals for a free-space medium
\begingroup
\begin{align}
I_{s,c} &{}= \int_{A_\test}\int_{A_\src} \frac{\cos(2\pi R)}{R}dA_\src dA_\test, \label{eq:cos}\\
I_{s,s} &{}= \int_{A_\test}\int_{A_\src} \frac{\sin(2\pi R)}{R}dA_\src dA_\test, \label{eq:sin}\\
I_{v,c} &{}= \int_{A_\test}(\mathbf{x}_\test-\mathbf{x}_j)\cdot\int_{A_\src} \frac{\cos(2\pi R)}{R}(\mathbf{x}_\src-\mathbf{x}_i)dA_\src dA_\test, \label{eq:cos_evp}\\
I_{v,s} &{}= \int_{A_\test}(\mathbf{x}_\test-\mathbf{x}_j)\cdot\int_{A_\src} \frac{\sin(2\pi R)}{R}(\mathbf{x}_\src-\mathbf{x}_i)dA_\src dA_\test, \label{eq:sin_evp}
\end{align}
\endgroup
where $\mathbf{x}_j=(1/20\,\mathrm{m},1/20\,\mathrm{m})$ and $\mathbf{x}_i=(-1/20\,\mathrm{m},1/20\,\mathrm{m})$.  When $\lambda=1$~m, \eqref{eq:cos} and~\eqref{eq:sin} respectively correspond to the even- and odd-term components of~\eqref{eq:esp2}, and~\eqref{eq:cos_evp} and~\eqref{eq:sin_evp} respectively correspond to the even- and odd-term components of~\eqref{eq:evp2}.  Additionally the maximum edge lengths of $A_\src$ and $A_\test$ are $1/10\,\lambda$.

\begin{table}
\centering
\begin{tabular}{c c c c c c c}
\toprule
Case & $\theta$ & $\Delta y$ & $\Delta z$ & Interaction & Potential \\
\midrule
1 & $\pz\pz0^\circ$ & 0          & 0          & Singular      & Scalar \\
2 & $  \pz45^\circ$ & 0          & 0          & Singular      & Scalar \\
3 & $  \pz90^\circ$ & 0          & 0          & Singular      & Scalar \\
4 & $    180^\circ$ & 0          & 0          & Singular      & Scalar \\
5 & $    180^\circ$ & 0          & $\delta_z$ & Near-singular & Scalar \\
6 & $\pz\pz0^\circ$ & $\delta_y$ & 0          & Far           & Scalar \\
7 & $  \pz90^\circ$ & 0          & 0          & Singular      & Vector \\
8 & $    180^\circ$ & 0          & 0          & Singular      & Vector \\
\bottomrule
\end{tabular}
\caption{Parameters describing the configurations of $A_\src$ and $A_\test$ to analyze singular, near-singular, and far interactions for scalar and vector potentials.  Note that Cases 4 and 8 are cases where the source and test triangles coincide.}
\label{tab:parameters}
\end{table}

The integrals $I_{s,c}$ and $I_{v,c}$ can be (nearly-)singular, depending on the distance between $A_\src$ and $A_\test$.  On the other hand, $I_{s,s}$ and $I_{v,s}$ are nonsingular.  Cases 1--4 and 7--8 in Table~\ref{tab:parameters}, with $\Delta y=\Delta z=0$, are singular, and Case 5, with $\Delta z=\delta_z$, is nearly singular.  In Case 6, $\Delta y=\delta_y$ is large enough that the integrands of $I_{s,c}$ and $I_{v,c}$ are smooth.

We compute reference solutions using Mathematica~\cite{mathematica} with 34 digits of working precision and precision and accuracy goals of 17 digits.  To compute $I_{s,c}$ and $I_{v,c}$, we use the radial--angular transformation presented in~\cite{khayat_2008}.  
To verify the implementation of the radial--angular transformation, we compute $I_{s,s}$ and $I_{v,s}$, which are nonsingular, with and without the transformation to confirm both reference solutions are the same.  
The amount of time required to compute each reference solution $I_{s,c}$ and $I_{v,c}$ is hours, compared to the fraction of a second required for the quadrature integration.

To perform the quadrature integration, which we denote by $\tilde{I}$, we use polynomial and polynomial-root Gauss--Legendre rules with the radial--angular transformation~\cite{khayat_2008} for the integral over $A_\src$.  For the one-dimensional polynomial rules, we use 100 points and, for the one-dimensional polynomial-root rules, we use 9 points.  For the integral over $A_\test$, we use two-dimensional polynomial rules~\cite{lyness_1975,dunavant_1985,wandzura_2003,papanicolopulos_2015}, the DMRW rules~\cite{vipiana_2013,ma_1996}, and the rules from Approaches 1 and 2 from Section~\ref{sec:quadrature}.  Because the DMRW rules are asymmetric, the points are concentrated at one of the vertices; therefore, there are three possibilities.  In these results, we compare the average error of these three DMRW choices.

\begin{figure}
\centering
\begin{tikzpicture}
\node at (0,0  ) {\includegraphics[scale=.65,clip=true,trim=2.35in 0.10in 2.87in 0.20in]{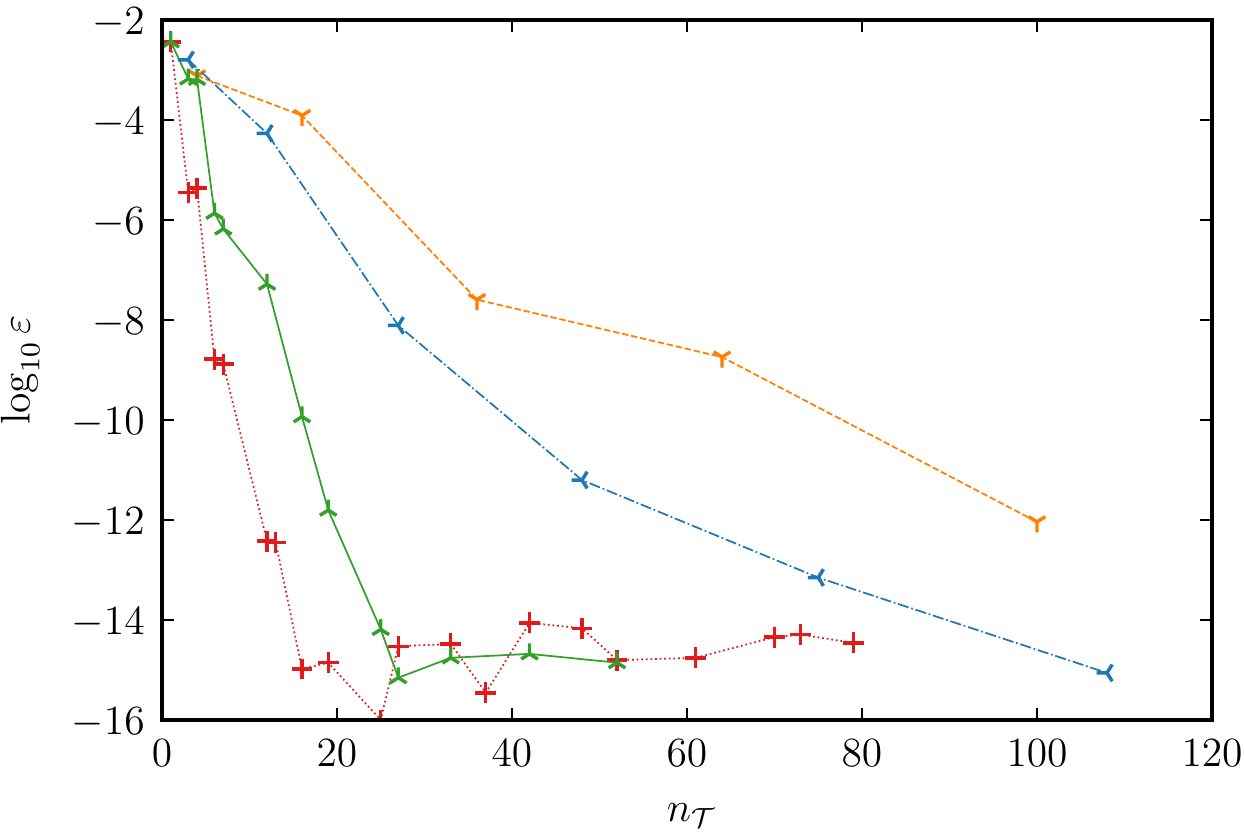}};
\node at (2.35,1.6) {\includegraphics[scale=.58,clip=true,trim=7.25in 1.4in 1.15in 1.10in]{t1_0_sin.pdf}};
\end{tikzpicture}
\vspace{-1em}
\caption{Relative error for $I_{s,s}$, corresponding to Case 1, with $\theta=0^\circ$, $\Delta y=0$, and $\Delta z=0$.  Similar convergence properties would be observed for the other cases listed in Table~\ref{tab:parameters}.}
\label{fig:t1_0_sin}
\end{figure}

Fig.~\ref{fig:t1_0_sin} shows the relative errors $\varepsilon=\left|(\tilde{I}-I)/I\right|$, with respect to the number of quadrature points used to compute the test integral, $\ntest$, for $I_{s,s}$ for Case 1, with $\theta=0^\circ$, $\Delta y=0$, and $\Delta z=0$.  
Because $I_{s,s}$ is not singular, the polynomial rules perform the most efficiently; however, Approach 1 achieves similar efficiency.  The averaged DMRW rules are the least efficient.  This trend also occurs for the other values of $\theta$; therefore, we omit those plots.  Furthermore, $I_{v,s}$ is also smooth and nonsingular, and shows similar properties to $I_{s,s}$.

\begin{figure}
\centering%
\begin{subfigure}[b]{0.5\columnwidth}
\flushleft
\begin{tikzpicture}
\node at (0,0  ) {\includegraphics[scale=.65,clip=true,trim=2.35in 0.10in 2.87in 0.20in]{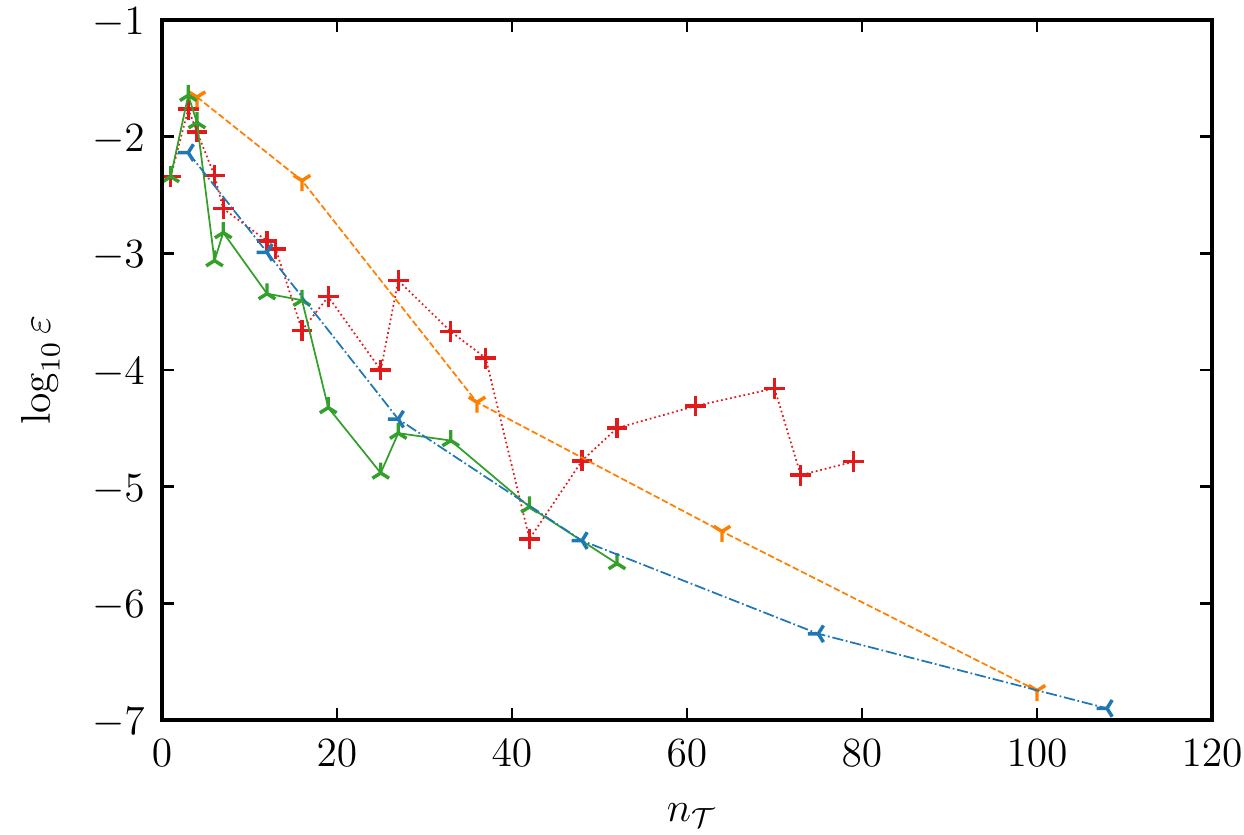}};
\end{tikzpicture}
\vspace{-1.75em}
\caption{$\theta=0^\circ$}
\label{fig:t1_0_cos}
\end{subfigure}%
\begin{subfigure}[b]{0.5\columnwidth}
\flushright
\begin{tikzpicture}
\node at (0,0  ) {\includegraphics[scale=.65,clip=true,trim=2.35in 0.10in 2.87in 0.20in]{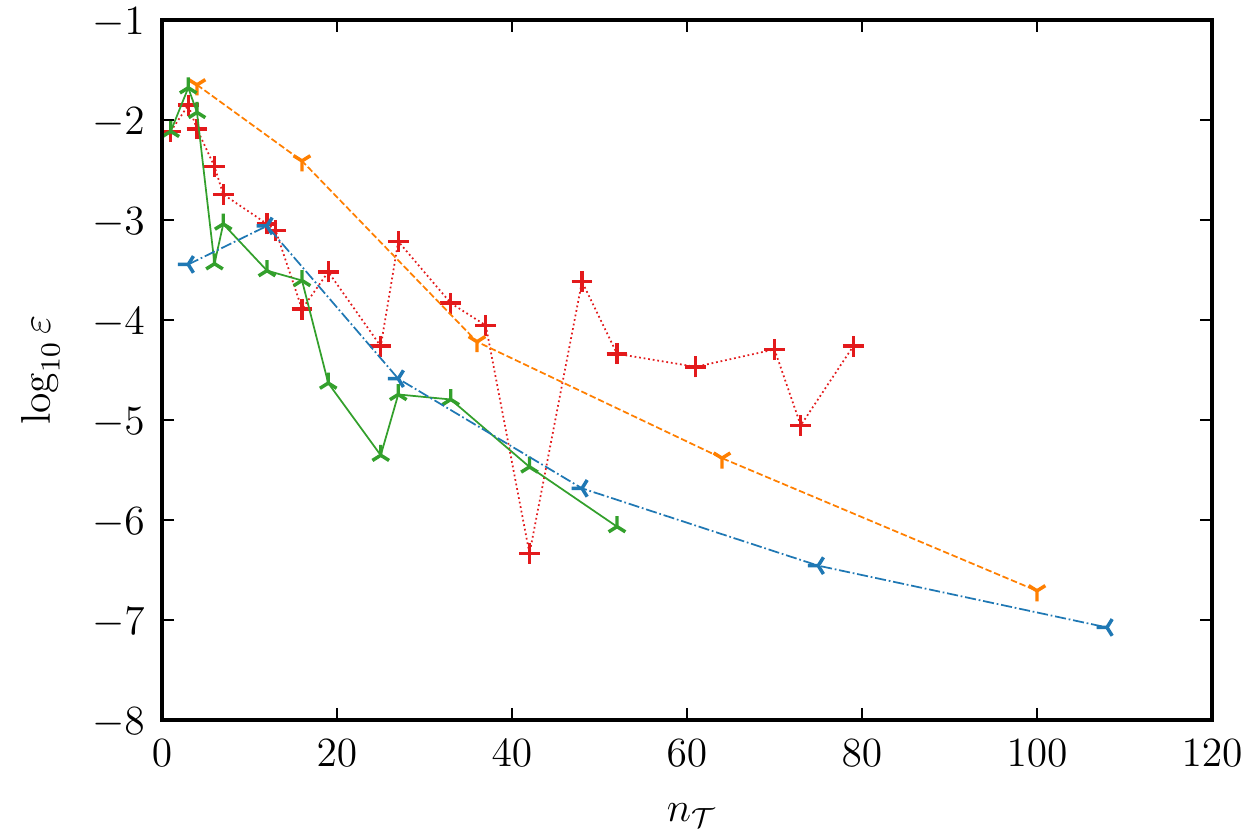}};
\node at (2.35,1.6) {\includegraphics[scale=.58,clip=true,trim=7.25in 1.4in 1.15in 1.10in]{t1_45_cos.pdf}};
\end{tikzpicture}
\vspace{-1.75em}
\caption{$\theta=45^\circ$}
\label{fig:t1_45_cos}
\end{subfigure}%
\\[1em]
\begin{subfigure}[b]{0.5\columnwidth}
\flushleft
\begin{tikzpicture}
\node at (0,0  ) {\includegraphics[scale=.65,clip=true,trim=2.35in 0.10in 2.87in 0.20in]{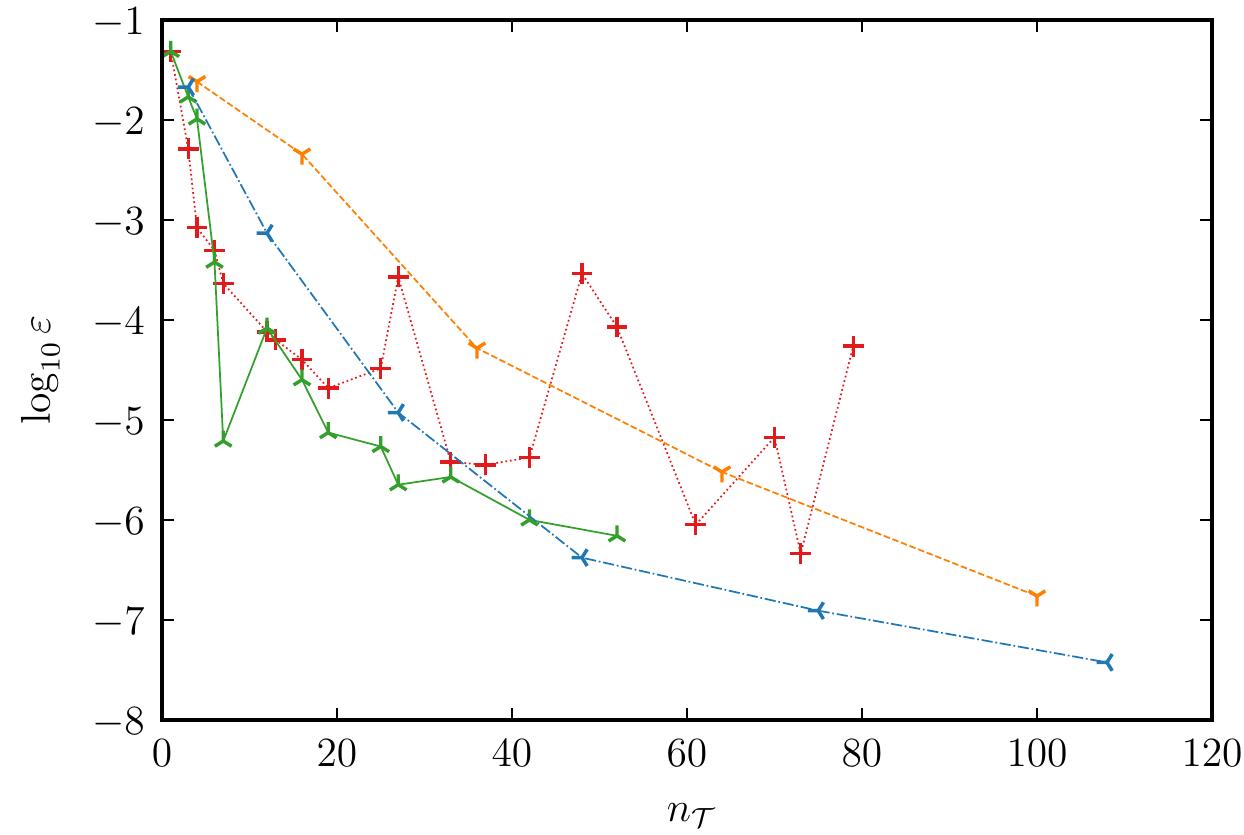}};
\end{tikzpicture}
\vspace{-1.75em}
\caption{$\theta=90^\circ$}
\label{fig:t1_90_cos}
\end{subfigure}%
\begin{subfigure}[b]{0.5\columnwidth}
\flushright
\begin{tikzpicture}
\node at (0,0  ) {\includegraphics[scale=.65,clip=true,trim=2.35in 0.10in 2.87in 0.20in]{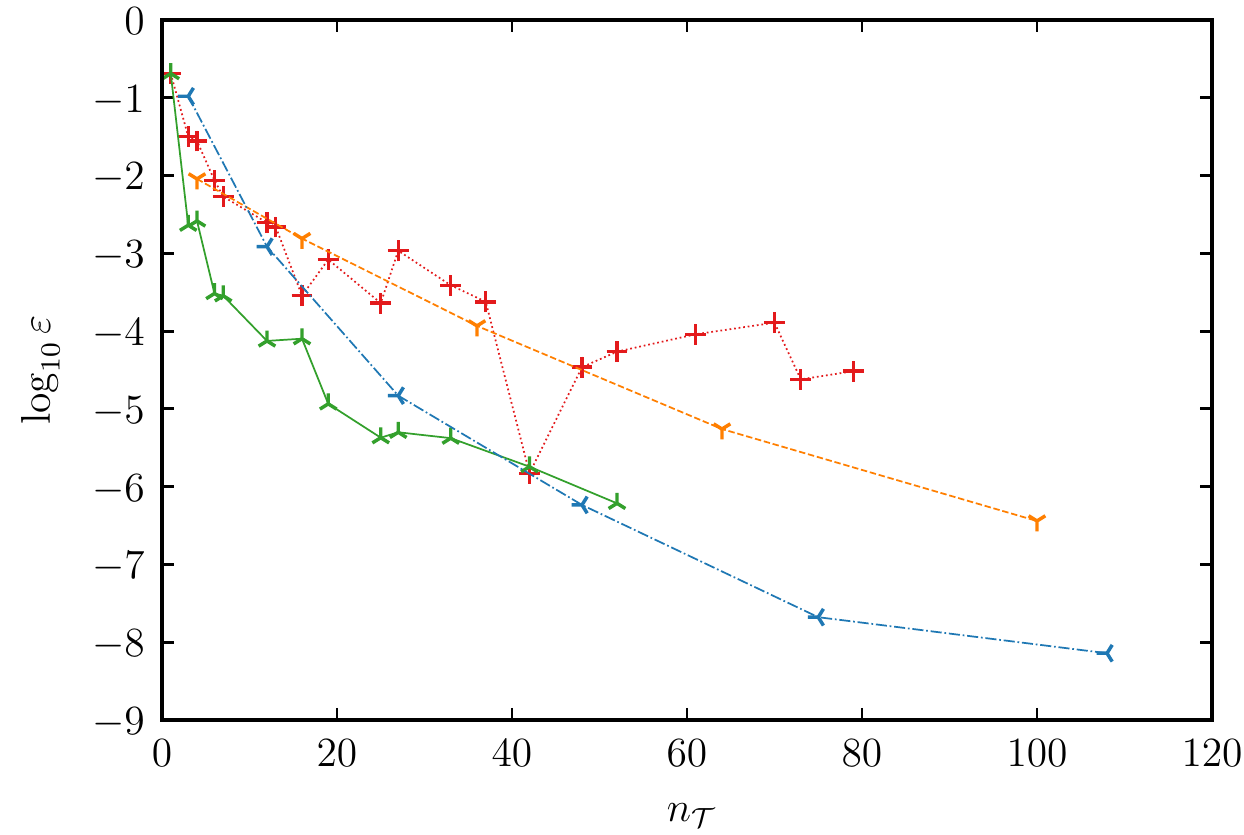}};
\end{tikzpicture}
\vspace{-1.75em}
\caption{$\theta=180^\circ$}
\label{fig:t1_180_cos}
\end{subfigure}
\caption{Relative error for $I_{s,c}$, corresponding to Cases 1--4, with $\Delta y=0$ and $\Delta z=0$.  Note that $\theta = 135^\circ$ has similar features to $\theta = 45^\circ$; thus, we omit that result.}
\label{fig:t1_0-180_cos}
\end{figure}

Fig.~\ref{fig:t1_0-180_cos} shows the relative errors for Cases 1--4, with $I_{s,c}$ for $\Delta y=0$ and $\Delta z=0$.  
For the coplanar cases ($\theta=0^\circ$ in Fig.~\ref{fig:t1_0_cos}) and ($\theta=180^\circ$ in Fig.~\ref{fig:t1_180_cos}), both approaches generally outperform the polynomial quadrature rules and the averaged DMRW rules, and Approach 1 often outperforms the polynomial quadrature rules by orders of magnitude.  In Fig.~\ref{fig:t1_180_cos}, for example, Approach 1 outperforms the polynomial rules by at least two orders of magnitude for $\npoints_\test=27$. 
For the noncoplanar cases ($\theta=45^\circ$ in Fig.~\ref{fig:t1_45_cos}) and ($\theta=90^\circ$ in Fig.~\ref{fig:t1_90_cos}), both approaches often outperform the polynomial quadrature rules and the averaged DMRW rules.  Approach 1, though designed for coplanar elements, generally outperforms the polynomial rules and the averaged DMRW rules.  Approach 2, though less efficient for small $\npoints_\test$, yields a monotonically decreasing relative error as the number of quadrature points is increased.
We believe that the singularities derived for the two special cases in Sections~\ref{sec:coplanar} and~\ref{sec:perpendicular} are more generally applicable, as the good convergence properties observed in Fig.~\ref{fig:t1_0-180_cos} suggest. 
The techniques shown in~\cite{wilton_1984} and~\cite{wilton_2019} may allow one to derive the singularities for completely general source and test triangle configurations.



\begin{figure}
\centering
\begin{tikzpicture}
\node at (0,0) {\includegraphics[scale=.65,clip=true,trim=2.35in 0.10in 2.87in 0.20in]{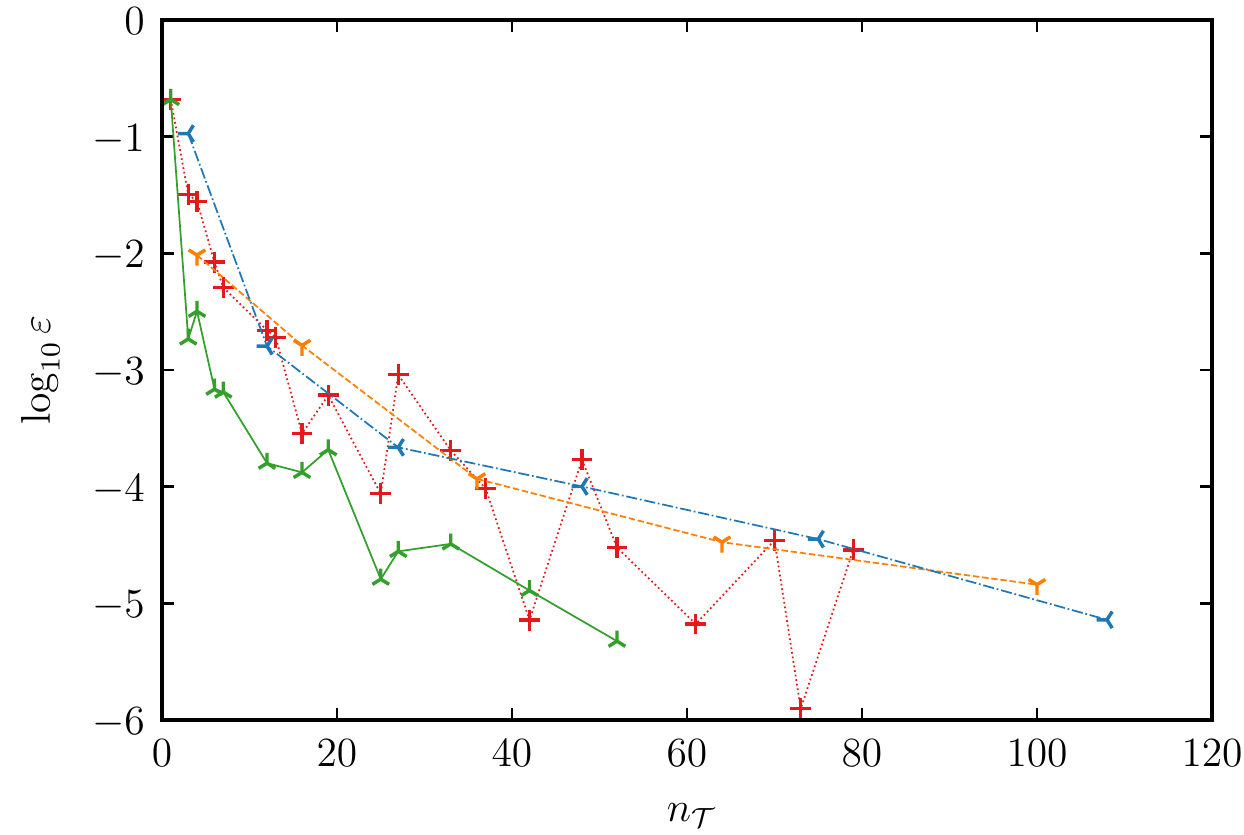}};
\node at (2.35,1.6) {\includegraphics[scale=.58,clip=true,trim=7.25in 1.4in 1.15in 1.10in]{t1_dz_180_cos.pdf}};
\end{tikzpicture}
\vspace{-1em}
\caption{Relative error for $I_{s,c}$, corresponding to Case 5, with $\theta=180^\circ$, $\Delta y=0$, and $\Delta z=\delta_z=1/2000$ m.}
\label{fig:t1_dz_180_cos}
\end{figure}

Fig.~\ref{fig:t1_dz_180_cos} shows the relative errors for $I_{s,c}$ for Case 5, with $\theta=180^\circ$, $\Delta y=0$, and $\Delta z=\delta_z$.  This case is nearly singular.  Approaches 1 and 2 are not designed for this case; nonetheless, Approach 1 often outperforms the polynomial rules and the averaged DMRW rules, and Approach 2 monotonically converges.

\begin{figure}
\centering
\begin{tikzpicture}
\node at (0,0) {\includegraphics[scale=.65,clip=true,trim=2.35in 0.10in 2.87in 0.20in]{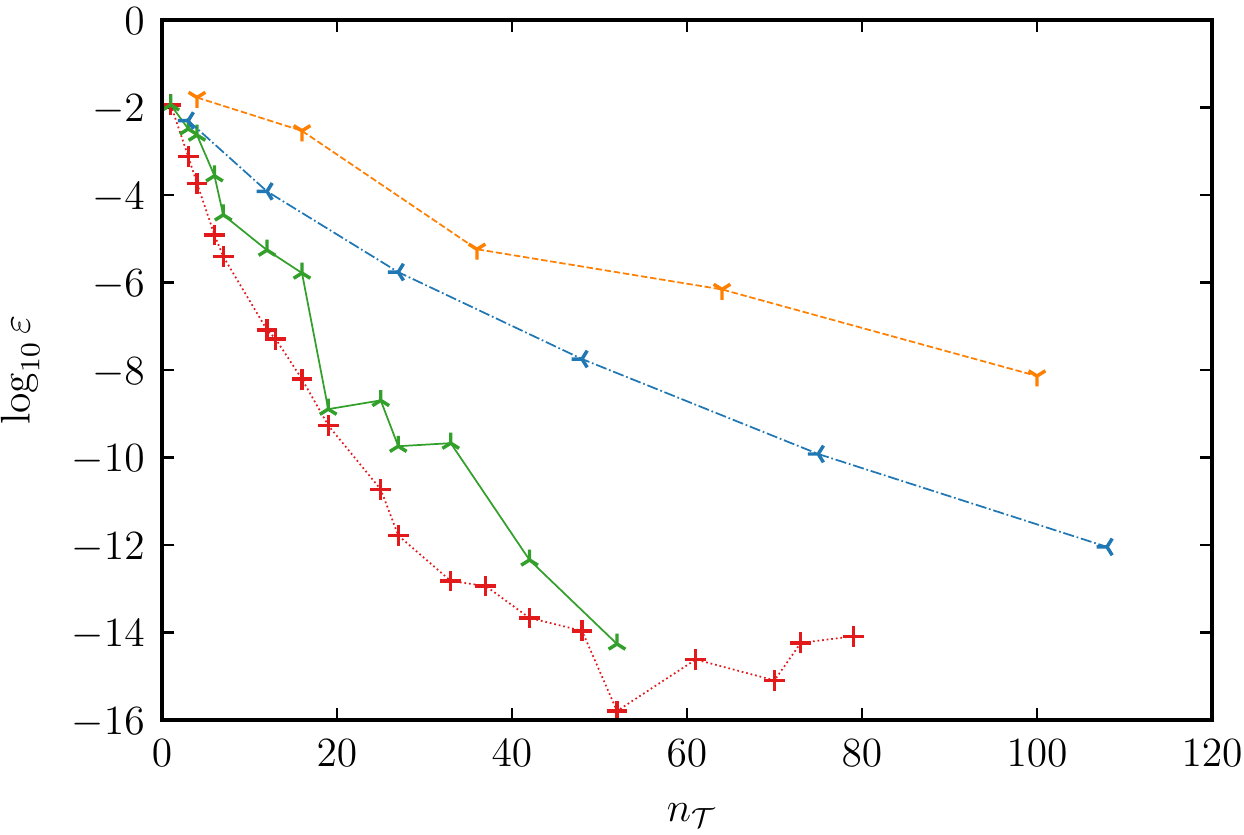}};
\node at (2.35,1.6) {\includegraphics[scale=.58,clip=true,trim=7.25in 1.4in 1.15in 1.10in]{t1_far_0_cos.pdf}};
\end{tikzpicture}
\vspace{-1em}
\caption{Relative error for $I_{s,c}$, corresponding to Case 6, with $\theta=0^\circ$, $\Delta y=\delta_y=(6\sqrt{2}-1)/60$ m, and $\Delta z=0$.}
\label{fig:t1_far_0_cos}
\end{figure}

Fig.~\ref{fig:t1_far_0_cos} shows the relative errors for $I_{s,c}$ for Case 6 with $\theta=0^\circ$, $\Delta y=\delta_y$, and $\Delta z=0$.  This case is not singular, but Approach 1 converges nearly as rapidly as the polynomial rules.  Approach 2 converges faster than the averaged DMRW rules.

\begin{figure*}
\begin{subfigure}[b]{0.5\columnwidth}
\flushleft
\begin{tikzpicture}
\node at (0,0  ) {\includegraphics[scale=.65,clip=true,trim=2.35in 0.10in 2.87in 0.20in]{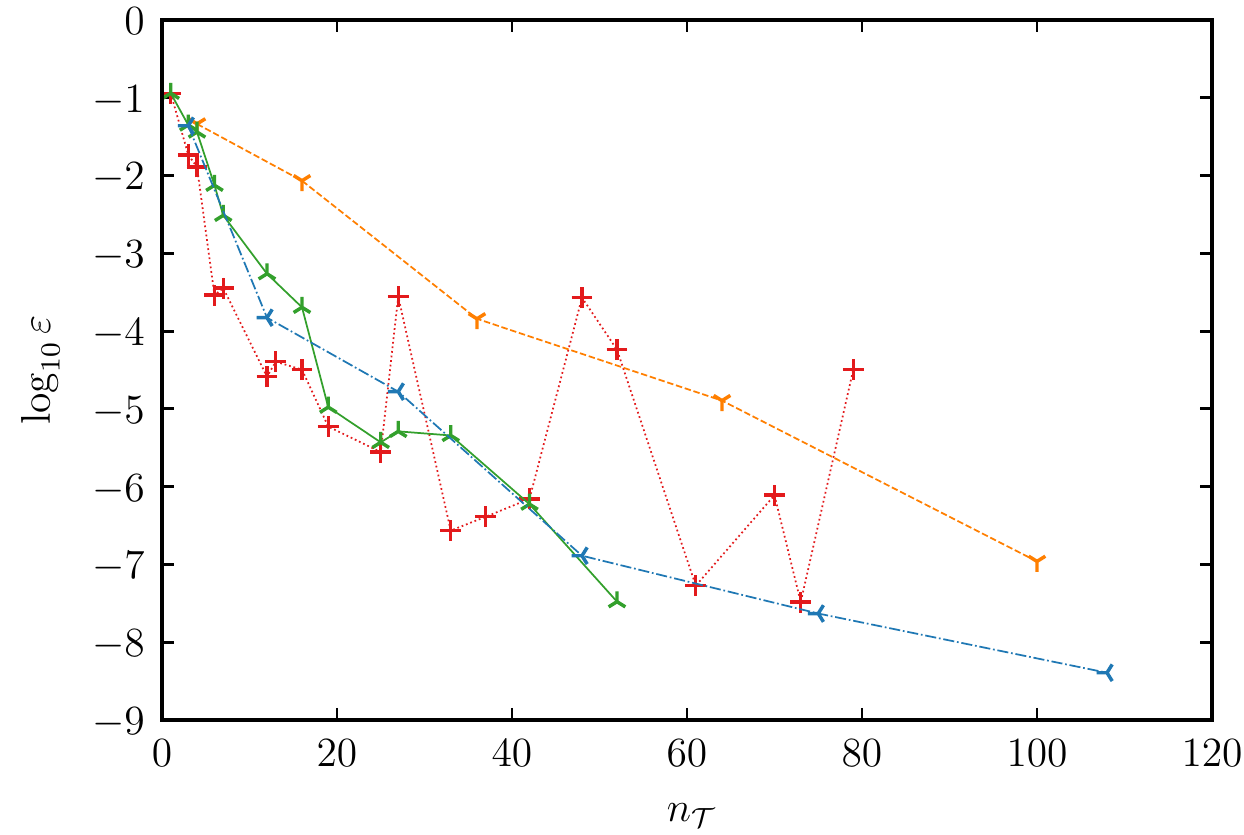}};
\end{tikzpicture}
\vspace{-1.75em}
\caption{$\theta=90^\circ$}
\label{fig:t1_evp_90_cos}
\end{subfigure}%
\begin{subfigure}[b]{0.5\columnwidth}
\flushright
\begin{tikzpicture}
\node at (0,0  ) {\includegraphics[scale=.65,clip=true,trim=2.35in 0.10in 2.87in 0.20in]{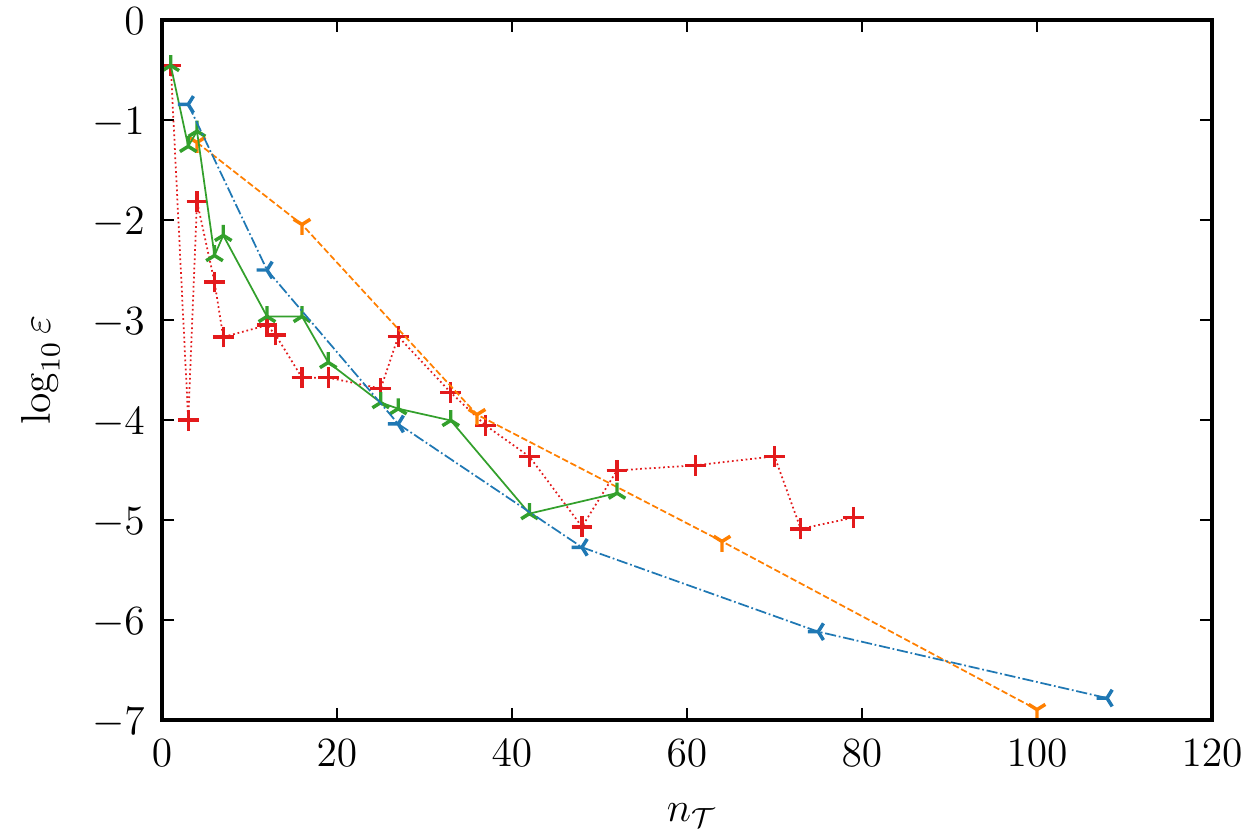}};
\node at (2.35,1.6) {\includegraphics[scale=.58,clip=true,trim=7.25in 1.4in 1.15in 1.10in]{t1_evp_180_cos.pdf}};
\end{tikzpicture}
\vspace{-1.75em}
\caption{$\theta=180^\circ$}
\label{fig:t1_evp_180_cos}
\end{subfigure}
\caption{Relative error for $I_{v,c}$, corresponding to Cases 7--8, with $\Delta y=0$ and $\Delta z=0$.}
\label{fig:t1_evp_90-180_cos}
\end{figure*}

Finally, Fig.~\ref{fig:t1_evp_90-180_cos} shows the relative errors for Cases 7--8, with $I_{v,c}$ for $\Delta y=0$ and $\Delta z=0$.  
For the coplanar case ($\theta=180^\circ$ in Fig.~\ref{fig:t1_evp_180_cos}), both approaches generally outperform the polynomial quadrature rules and the averaged DMRW rules for larger numbers of points. 
For the noncoplanar case ($\theta=90^\circ$ in Fig.~\ref{fig:t1_evp_90_cos}), Approach 2 and the averaged DMRW rules yield monotonically decreasing relative errors as the number of quadrature points is increased, and Approach 1 fluctuates less than the polynomial quadrature rules.

%% file: fig_two_triangles.tex
\newsavebox{\twotriangles}%
\sbox{\twotriangles}{%
\newcommand*\projA{-20}
\newcommand*\projB{ 35}
%
\definecolor{darkred} {RGB}{227,26,28}%
\definecolor{darkorange} {RGB}{255,127,0}%
\definecolor{lightred} {RGB}{243.8,163.4,164.2}%
\definecolor{darkblue}  {RGB}{ 31,120,180}%
\definecolor{lightblue}  {RGB}{165.4,201,225}
\definecolor{darkgreen} {RGB}{ 51,160, 44}%
\definecolor{lightgreen}{RGB}{173.4,217,170.6}
\def\xangle{100}
\def\yangle{60}
\begin{tikzpicture}[scale=0.6,
x={({ sin(\xangle)*1cm},{ cos(\xangle)*1cm})}, 
y={({.9*sin(\yangle)*1cm},{.9*cos(\yangle)*1cm})}, 
z={(0cm,1cm)}
    ]   

   \def\Deltaz{1};    
   \def\Deltay{3};    
   \def\h{4};         
   \def\xl{-4};       
   \def\xr{4};        
   \def\meas{1};      
   \def\mw{.25};      
   \def\thetaval{45}; 
   \def\arcrad{1};    
   \def\ep{.1}        
   \def\ymax{13}      
   
   \useasboundingbox (-2cm,-2cm) rectangle (13cm,10.5cm);
   
   \begin{scope}[canvas is xy plane at z=0,transform shape]
   \draw[line cap=round] (0,0) -- (8-\ep, 0);
   \draw[line cap=round] (0,0) -- (0,\ymax-\ep);
   \pgflowlevelsynccm
   \draw[-{Stealth[scale=2.0]},line cap=round] (0,0) to (8, 0) node[right,scale=2,dash pattern=on 0cm off 100cm] {$x$};
   \draw[-{Stealth[scale=2.0]},line cap=round] (0,0) to (0,\ymax) node[above,scale=2,dash pattern=on 0cm off 100cm] {$y$};
   \end{scope}

   \begin{scope}[canvas is xz plane at y=0]
   \draw[line cap=round] (0,0) -- (0,8-\ep);
   \pgflowlevelsynccm
   \draw[-{Stealth[scale=2.0]},line cap=round,dash pattern=on 0cm off 100cm] (0,0) to (0,8) node[above,scale=2] {$z$};
   \end{scope}

   \begin{scope}[canvas is xy plane at z=0]
      \draw[thick,darkblue,fill=lightblue,fill opacity=.5,draw opacity=1] (0,0) -- (\xr,\h) -- (\xl,\h) -- cycle;     
      \node[transform shape,scale=2,text=darkblue,right] at (0,\h/3*2) {$A_\src$};

      \draw[draw=darkblue,fill=darkblue] (\xl,\h) circle (.1);
      \node[transform shape,scale=2,text=darkblue,anchor=south] at (\xl-.0,\h-0) {$\mathbf{x}_i$};
      
      \draw[draw=darkblue,fill=darkblue] (\xl/4,\h*.7) circle (.1);
      \node[transform shape,scale=2,text=darkblue,anchor=north] at (\xl/4,\h*.7-.1) {$\mathbf{x}_\src$};
   \end{scope}

   \begin{scope}[canvas is xy plane at z=0,transform shape]
      \draw[line cap=round,gray] (\xr+\meas,\h+\ep) -- (\xr+\meas,\h+\Deltay-\ep); 
      \draw[line cap=round,dashed,gray] (\xr,\h) -- (\xr+\meas+\mw,\h); 
      \draw[line cap=round,dashed,gray] (\xr,\h+\Deltay) -- (\xr+\meas+\mw,\h+\Deltay); 
      \pgflowlevelsynccm
      \draw[{Stealth[scale=2.0,fill=gray,gray]}-{Stealth[scale=2.0,fill=gray,gray]},line cap=round,dash pattern=on 0cm off 100cm] (\xr+\meas,\h) to(\xr+\meas,\h+\Deltay);
      \node[transform shape,scale=2,right,gray] at (\xr+\meas+\mw,\h+\Deltay/2) {$\Delta y$};
   \end{scope}

   \begin{scope}[canvas is xz plane at y=\h+\Deltay]
      \draw[line cap=round,gray] (\xr+\meas,\ep) -- (\xr+\meas,\Deltaz-\ep); 
      \draw[line cap=round,dashed,gray] (\xr,\Deltaz) -- (\xr+\meas+\mw,\Deltaz); 
      \pgflowlevelsynccm
      \draw[{Stealth[scale=2.0,fill=gray,gray]}-{Stealth[scale=2.0,fill=gray,gray]},line cap=round,dash pattern=on 0cm off 100cm] (\xr+\meas,0) to (\xr+\meas,\Deltaz);
      \node[transform shape,scale=2,right,gray] at (\xr+\meas+\mw,\Deltaz/2) {$\Delta z$};
   \end{scope} 
   
   \begin{scope}[canvas is xy plane at z=\Deltaz,transform shape]
      \draw[dashed,line cap=round,gray] (0,\h+\Deltay) -- (0,\ymax-\ep);

      \pgflowlevelsynccm
      \draw[-{Stealth[scale=2.0,fill=gray,gray]},line cap=round,dash pattern=on 0cm off 100cm,text=gray] (0,\h+\Deltay) to (0,\ymax) node[above,scale=2] {$y'$};
   \end{scope}

   \begin{scope}[canvas is yz plane at x=0,transform shape]
      \draw[line cap=round,gray] (\h+\Deltay+\arcrad,\Deltaz) arc (0:\thetaval:\arcrad);
      \node[scale=2] at ({\h+\Deltay+1.5*\arcrad*cos(\thetaval/2)},{\Deltaz+1.5*\arcrad*sin(\thetaval/2)}) {$\theta$};
      \pgflowlevelsynccm
      
   \end{scope}

   \begin{scope}[canvas is plane={O(0,\h+\Deltay,\Deltaz)x(1,\h+\Deltay,\Deltaz)y(0,{\h+\Deltay+cos(\thetaval)},{\Deltaz+sin(\thetaval)})}]
      \draw[line cap=round,dashed,gray] (0,0) -- (0,\ymax-\h-\Deltay-\ep);
      
      \draw[thick,draw=darkgreen,fill=lightgreen,fill opacity=.5,draw opacity=1] (\xr,0) -- (0,\h) -- (\xl,0) -- cycle;
      
      \node[transform shape,scale=2,text=darkgreen,left] at (0,\h/3) {$A_\test$};
      
      \pgflowlevelsynccm
      \draw[-{Stealth[scale=2.0,fill=gray,gray]},line cap=round,dash pattern=on 0cm off 100cm,darkgreen] (0,0) to (0,1.5*\h); 
      
      \draw[draw=darkgreen,fill=darkgreen] (\xr,0) circle (.1);
      \node[transform shape,scale=2,text=darkgreen,anchor=south west] at (\xr-.2,0-.2) {$\mathbf{x}_j$};
      
      \draw[draw=darkgreen,fill=darkgreen] (\xl*.8,\h*.1) circle (.1);
      \node[transform shape,scale=2,text=darkgreen,anchor=south west] at (\xl*.8,\h*.1-.1) {$\mathbf{x}_\test$};
      
   \end{scope}
   
   \def\R{sqrt((\xl*.8-\xl/4)^2+(\h+\Deltay+\h*.1*cos(\thetaval)-\h*.7)^2+(\Deltaz+\h*.1*sin(\thetaval))^2)};
   \draw[line cap=round,gray] ({\xl/4+(\xl*.8-\xl/4)/\R*\ep},{\h*.7+(\h+\Deltay+\h*.1*cos(\thetaval)-\h*.7)/\R*\ep},{0+(\Deltaz+\h*.1*sin(\thetaval))/\R*\ep}) -- ({\xl*.8-(\xl*.8-\xl/4)/\R*\ep}, {\h+\Deltay+\h*.1*cos(\thetaval)-(\h+\Deltay+\h*.1*cos(\thetaval)-\h*.7)/\R*\ep}, {\Deltaz+\h*.1*sin(\thetaval)-(\Deltaz+\h*.1*sin(\thetaval))/\R*\ep}) node[midway,right,gray] {$R$};

   \draw[{Stealth[scale=2.0,fill=gray,gray]}-{Stealth[scale=2.0,fill=gray,gray]},line cap=round,dash pattern=on 0cm off 100cm] (\xl/4,\h*.7,0) to (\xl*.8, {\h+\Deltay+\h*.1*cos(\thetaval)}, {\Deltaz+\h*.1*sin(\thetaval)});

\end{tikzpicture}%
}%

%% file: conclusions.tex
\section{Conclusions}
\label{sec:conclusions}

In this paper, we characterized the logarithmic singularities encountered in the method of moments for the EFIE, and we used two approaches for designing geometrically symmetric quadrature rules to integrate these singular integrands.
Our first approach was often able to outperform polynomial rules by several orders of magnitude for singular cases and exhibited similar convergence properties for nonsingular cases.
Though not as efficient as Approach 1 for singular integrals or the polynomial rules for nonsingular integrals, the relative error arising from Approach 2 decreases monotonically with respect to the number of integration points, a feature that is not observed with the polynomial scheme when applied to integrands with logarithmic singularities.  Additionally, for large numbers of integration points, the points arising from Approach 2 take less time to compute than those from Approach 1 since they are computed from one-dimensional rules.  Generally, Approaches 1 and 2, which have the favorable property of geometric symmetry, outperform the averaged DMRW rules.


%% file: acknowledgments.tex
\section*{Acknowledgments}

This paper describes objective technical results and analysis. Any subjective views or opinions that might be expressed in the paper do not necessarily represent the views of the U.S. Department of Energy or the United States Government.
Sandia National Laboratories is a multimission laboratory managed and operated by National Technology and Engineering Solutions of Sandia, LLC., a wholly owned subsidiary of Honeywell International, Inc., for the U.S. Department of Energy's National Nuclear Security Administration under contract DE-NA-0003525.

%% file: appendix_a.tex

\section{Quadrature Points and Weights for Approach 1}
\label{app:a}

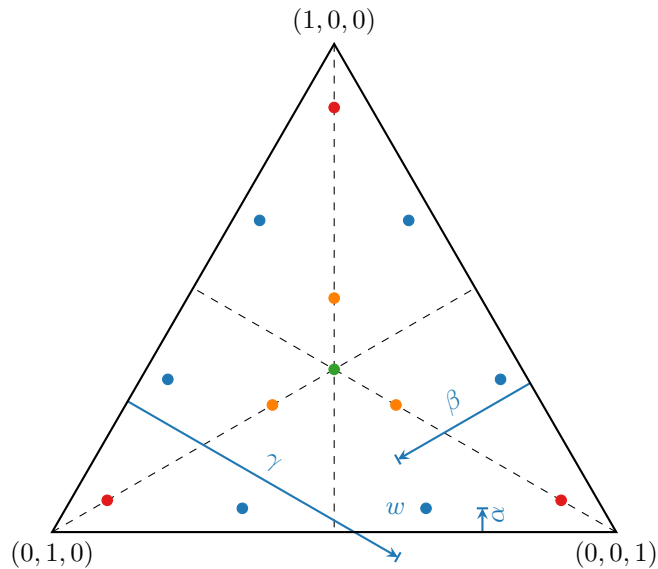
\begin{figure}[htbp!]
\vspace*{.5em}
\centering
\input{opt_fig_ieee.tex}
\vspace{-1em}
\caption{Pictorial representation of Approach 1 for a (1,2,1) rule.  Note each set of colored points is associated with a single orbit and is therefore symmetric.}
\label{fig:approach_1}
\end{figure}

Table~\ref{tab:approach_1} 
provides
the points and weights for Approach 1, which have been ordered similarly to those in~\cite{dunavant_1985} to facilitate comparison.  Fig.~\ref{fig:approach_1} shows a pictorial representation of Approach 1 for a (1,2,1) rule, illustrating $w$, $\alpha$, $\beta$, and $\gamma$ (more details in~\cite{freno_quad}).

\begin{table}[htbp!]
\centering
\scalebox{.7}{%
\begin{tabular}{@{} c @{\hspace{1em}} c @{\hspace{1em}} c @{\hspace{1em}} c @{\hspace{1em}} c @{}}
\toprule
$n$  & $w$ & $\alpha$ & $\beta$ & $\gamma$ \\
\midrule
$1$    & $\ps1.000000000000000$ & $0.333333333333333$ & $0.333333333333333$ & $0.333333333333333$ \\ \midrule
$3$    & $\ps0.333333333333333$ & $0.695378779571022$ & $0.152310610214489$ & $0.152310610214489$ \\ \midrule
$4$    & $  -0.714433957991885$ & $0.333333333333333$ & $0.333333333333333$ & $0.333333333333333$ \\
       & $\ps0.571477985997295$ & $0.609421762429183$ & $0.195289118785409$ & $0.195289118785409$ \\ \midrule
$6$    & $\ps0.257014376989061$ & $0.097813388052768$ & $0.451093305973616$ & $0.451093305973616$ \\
       & $\ps0.076318956344273$ & $0.879676863242546$ & $0.060161568378727$ & $0.060161568378727$ \\ \midrule   
$7$    & $\ps0.285195062745114$ & $0.333333333333333$ & $0.333333333333333$ & $0.333333333333333$ \\
       & $\ps0.069920501732796$ & $0.005055144001731$ & $0.497472427999135$ & $0.497472427999135$ \\
       & $\ps0.168347810685499$ & $0.751954885659122$ & $0.124022557170439$ & $0.124022557170439$ \\ \midrule   
$12$   & $\ps0.149245346570655$ & $0.516608648914836$ & $0.241695675542582$ & $0.241695675542582$ \\
       & $\ps0.069354579325139$ & $0.841888071019192$ & $0.079055964490404$ & $0.079055964490404$ \\
       & $\ps0.057366703718770$ & $0.027954810349577$ & $0.337219412523235$ & $0.634825777127188$ \\ \midrule   
$16$   & $\ps0.118284309157793$ & $0.333333333333333$ & $0.333333333333333$ & $0.333333333333333$ \\
       & $\ps0.124676966089597$ & $0.097302416356286$ & $0.451348791821857$ & $0.451348791821857$ \\
       & $\ps0.074683120349995$ & $0.673868971638020$ & $0.163065514180990$ & $0.163065514180990$ \\
       & $\ps0.007304121966484$ & $0.990989752987059$ & $0.004505123506470$ & $0.004505123506470$ \\
       & $\ps0.043620510937330$ & $0.021026075771245$ & $0.199787336300406$ & $0.779186587928348$ \\ \midrule    
$19$   & $\ps0.101432563802204$ & $0.333333333333333$ & $0.333333333333333$ & $0.333333333333333$ \\
       & $\ps0.017018388271078$ & $0.000125645563818$ & $0.499937177218091$ & $0.499937177218091$ \\
       & $\ps0.082018035712029$ & $0.101603006817710$ & $0.449198496591145$ & $0.449198496591145$ \\
       & $\ps0.084388621611840$ & $0.602664000744707$ & $0.198667999627647$ & $0.198667999627647$ \\
       & $\ps0.026463125157577$ & $0.907319860390893$ & $0.046340069804553$ & $0.046340069804553$ \\
       & $\ps0.044817153990037$ & $0.038417534639478$ & $0.220337640197156$ & $0.741244825163365$ \\ \midrule       
$25$   & $\ps0.105091420511953$ & $0.333333333333333$ & $0.333333333333333$ & $0.333333333333333$ \\
       & $\ps0.037661425819142$ & $0.026375407117687$ & $0.486812296441157$ & $0.486812296441157$ \\
       & $\ps0.042086789369811$ & $0.817865014875771$ & $0.091067492562114$ & $0.091067492562114$ \\
       & $\ps0.078786977430742$ & $0.132651474581117$ & $0.299355462925693$ & $0.567993062493189$ \\
       & $\ps0.025634091780525$ & $0.020973903258140$ & $0.225844604082180$ & $0.753181492659680$ \\
       & $\ps0.004856253108931$ & $0.000612864423452$ & $0.048052095134699$ & $0.951335040441849$ \\ \midrule 
$27$   & $\ps0.013922250132469$ & $0.001868409018971$ & $0.499065795490515$ & $0.499065795490515$ \\
       & $\ps0.057067640806032$ & $0.126286102659602$ & $0.436856948670199$ & $0.436856948670199$ \\
       & $\ps0.075852683253245$ & $0.482498683485778$ & $0.258750658257111$ & $0.258750658257111$ \\
       & $\ps0.054618615030858$ & $0.721464823901483$ & $0.139267588049258$ & $0.139267588049258$ \\
       & $\ps0.017391893221775$ & $0.923513301673739$ & $0.038243349163131$ & $0.038243349163131$ \\
       & $\ps0.041665767115182$ & $0.052947102900912$ & $0.309884397819744$ & $0.637168499279344$ \\
       & $\ps0.015574358329295$ & $0.015157675671346$ & $0.166657071430478$ & $0.818185252898176$ \\ \midrule        
$33$   & $\ps0.010658830647119$ & $0.000961899381510$ & $0.499519050309245$ & $0.499519050309245$ \\
       & $\ps0.025956997773006$ & $0.051276629861635$ & $0.474361685069183$ & $0.474361685069183$ \\
       & $\ps0.065499752081693$ & $0.462737337945776$ & $0.268631331027112$ & $0.268631331027112$ \\
       & $\ps0.048946628533910$ & $0.744253274521589$ & $0.127873362739205$ & $0.127873362739205$ \\
       & $\ps0.014630278533084$ & $0.927767586967484$ & $0.036116206516258$ & $0.036116206516258$ \\
       & $\ps0.027634085586402$ & $0.030395375318387$ & $0.256349252474986$ & $0.713255372206627$ \\
       & $\ps0.048368260533086$ & $0.327268158517310$ & $0.122017720527821$ & $0.550714120954869$ \\ 
       & $\ps0.007818076762773$ & $0.010598772295753$ & $0.129086525733381$ & $0.860314701970866$ \\\midrule  
$42$   & $\ps0.024915420246230$ & $0.024301132858219$ & $0.487849433570890$ & $0.487849433570890$ \\
       & $\ps0.047276822487624$ & $0.137728483927436$ & $0.431135758036282$ & $0.431135758036282$ \\
       & $\ps0.059863248782438$ & $0.459456507967625$ & $0.270271746016188$ & $0.270271746016188$ \\
       & $\ps0.040170006809921$ & $0.654636150686281$ & $0.172681924656860$ & $0.172681924656860$ \\
       & $\ps0.006946885444244$ & $0.801965072360505$ & $0.099017463819748$ & $0.099017463819748$ \\
       & $\ps0.008653587757130$ & $0.947249960628273$ & $0.026375019685864$ & $0.026375019685864$ \\
       & $\ps0.021475651931304$ & $0.044210014484261$ & $0.140466826963294$ & $0.815323158552444$ \\
       & $\ps0.036759184755577$ & $0.079669538941889$ & $0.299092091530278$ & $0.621238369527833$ \\
       & $\ps0.011484252241907$ & $0.008862194513821$ & $0.291941058549882$ & $0.699196746936297$ \\
       & $\ps0.003034591974085$ & $0.000363031589062$ & $0.112034280606847$ & $0.887602687804090$ \\ \midrule     
$52$   & $\ps0.035913703327725$ & $0.333333333333333$ & $0.333333333333333$ & $0.333333333333333$ \\
       & $\ps0.007684616702762$ & $0.006489122618348$ & $0.496755438690826$ & $0.496755438690826$ \\
       & $\ps0.036063012009317$ & $0.190402198316418$ & $0.404798900841791$ & $0.404798900841791$ \\
       & $\ps0.028354991546057$ & $0.088553788863145$ & $0.455723105568427$ & $0.455723105568427$ \\
       & $\ps0.038915008624662$ & $0.507526417497225$ & $0.246236791251388$ & $0.246236791251388$ \\
       & $\ps0.026246895702946$ & $0.697435436357295$ & $0.151282281821353$ & $0.151282281821353$ \\
       & $\ps0.014915840667527$ & $0.839778294877712$ & $0.080110852561144$ & $0.080110852561144$ \\
       & $\ps0.004420380087181$ & $0.962899004829267$ & $0.018550497585367$ & $0.018550497585367$ \\
       & $\ps0.030617247512664$ & $0.114694286356458$ & $0.283573119736518$ & $0.601732593907024$ \\
       & $\ps0.019595437900154$ & $0.031771068006969$ & $0.362387192929587$ & $0.605841739063444$ \\
       & $\ps0.005257310837977$ & $0.001983209875968$ & $0.248665339750671$ & $0.749351450373361$ \\
       & $\ps0.018884511581608$ & $0.046839465739992$ & $0.192752718293175$ & $0.760407815966833$ \\ 
       & $\ps0.008026168942750$ & $0.014762645440057$ & $0.094540889252372$ & $0.890696465307571$ \\    
\bottomrule
\end{tabular}
}
\caption{Quadrature points and weights for Approach 1.}
\label{tab:approach_1}
\end{table}

\FloatBarrier

%% file: opt_fig_ieee.tex
\begin{tikzpicture}

\definecolor{lightgreen}{RGB}{173.4,217,170.6}
\definecolor{darkgreen} {RGB}{227  ,26,28}
\definecolor{darkblue}  {RGB}{ 31  ,120,180}
\definecolor{lightblue} {RGB}{165.4,201,225}
\definecolor{darkred}   {RGB}{ 51  ,160, 44}
\definecolor{lightred}  {RGB}{243.8,163.4,164.2}
\definecolor{darkorange}{RGB}{255  ,127  ,0}

\def\ts{7.5}; 

\coordinate (A) at (-0.5,{-sqrt(3)/6});
\coordinate (B) at ( 0.5,{-sqrt(3)/6});
\coordinate (C) at ( 0  , {sqrt(1/3)});
\coordinate (D) at ($0.5*(A)+0.5*(B)$);
\coordinate (E) at ($0.5*(B)+0.5*(C)$);
\coordinate (F) at ($0.5*(C)+0.5*(A)$);
\coordinate (O) at (0,0);

\def\tx{0}
\def\ty{{\ts*sqrt(3)/6}}

\coordinate (T1) at ($\ts*(A)+(\tx,\ty)$);
\coordinate (T2) at ($\ts*(B)+(\tx,\ty)$);
\coordinate (T3) at ($\ts*(C)+(\tx,\ty)$);
\coordinate (T4) at ($\ts*(D)+(\tx,\ty)$);

\node[above = 0 of T3] {$(1,0,0)$};
\node[anchor=north] at (T2) {$(0,0,1)$};
\node[anchor=north] at (T1) {$(0,1,0)$};

\draw[dashed] ($\ts*(C)+(\tx,\ty)$) -- ($\ts*(D)+(\tx,\ty)$);
\draw[dashed] ($\ts*(A)+(\tx,\ty)$) -- ($\ts*(E)+(\tx,\ty)$);
\draw[dashed] ($\ts*(B)+(\tx,\ty)$) -- ($\ts*(F)+(\tx,\ty)$);

\draw[draw=darkred,fill=darkred] ($\ts*(  0.00000000000000,  0.00000000000000)+(\tx,\ty)$) circle (.07);
\draw[draw=darkorange,fill=darkorange] ($\ts*( -0.10948105088144, -0.06320891419756)+(\tx,\ty)$) circle (.07);
\draw[draw=darkorange,fill=darkorange] ($\ts*(  0.10948105088144, -0.06320891419756)+(\tx,\ty)$) circle (.07);
\draw[draw=darkorange,fill=darkorange] ($\ts*(  0.00000000000000,  0.12641782839512)+(\tx,\ty)$) circle (.07);
\draw[draw=darkgreen,fill=darkgreen] ($\ts*( -0.40230484564668, -0.23227081093040)+(\tx,\ty)$) circle (.07);
\draw[draw=darkgreen,fill=darkgreen] ($\ts*(  0.40230484564668, -0.23227081093040)+(\tx,\ty)$) circle (.07);
\draw[draw=darkgreen,fill=darkgreen] ($\ts*(  0.00000000000000,  0.46454162186080)+(\tx,\ty)$) circle (.07);
\draw[draw=darkblue,fill=darkblue] ($\ts*(  0.13208759028978,  0.26423375160519)+(\tx,\ty)$) circle (.07); 
\draw[draw=darkblue,fill=darkblue] ($\ts*(  0.29487693657225, -0.01772566708697)+(\tx,\ty)$) circle (.07); 
\draw[draw=darkblue,fill=darkblue] ($\ts*( -0.13208759028978,  0.26423375160519)+(\tx,\ty)$) circle (.07); 
\draw[draw=darkblue,fill=darkblue] ($\ts*(  0.16278934628247, -0.24650808451821)+(\tx,\ty)$) circle (.07); 
\draw[draw=darkblue,fill=darkblue] ($\ts*( -0.29487693657225, -0.01772566708697)+(\tx,\ty)$) circle (.07); 
\draw[draw=darkblue,fill=darkblue] ($\ts*( -0.16278934628247, -0.24650808451821)+(\tx,\ty)$) circle (.07); 


%
%
%
%

\draw[thick,draw=darkblue,fill=darkblue,->,>=stealth] ($\ts*( 0.16278934628247+.10,  {-sqrt(3)/6})+(\tx,\ty)$) -- ($\ts*(  0.16278934628247+.10,  -0.24650808451821)+(\tx,\ty)$) node [darkblue,near end, below,rotate=90] {$\alpha$};
\draw[thick,draw=darkblue,fill=darkblue] ($\ts*(0.16278934628247+.09,  -0.24650808451821)+(\tx,\ty)$) -- ($\ts*(0.16278934628247+.11,  -0.24650808451821)+(\tx,\ty)$);

\draw[thick,draw=darkblue,fill=darkblue,->,>=stealth] 
($\ts*( 0.16278934628247, -0.24650808451821)+\ts*.10*(-.5,{sqrt(3)/2})+\ts*0.270949467507843*({sqrt(3)/2},.5)+(\tx,\ty)$) -- 
($\ts*( 0.16278934628247, -0.24650808451821)+\ts*.10*(-.5,{sqrt(3)/2})+(\tx,\ty)$) node [darkblue,midway, above,rotate=30] {$\beta$}; 
\draw[thick,draw=darkblue,fill=darkblue] 
($\ts*( 0.16278934628247, -0.24650808451821)+\ts*.11*(-.5,{sqrt(3)/2})+(\tx,\ty)$) -- 
($\ts*( 0.16278934628247, -0.24650808451821)+\ts*.09*(-.5,{sqrt(3)/2})+(\tx,\ty)$);

\draw[thick,draw=darkblue,fill=darkblue,->,>=stealth] 
($\ts*( 0.16278934628247, -0.24650808451821)+\ts*.10*(-.5,-{sqrt(3)/2})+\ts*0.552908886200003*(-{sqrt(3)/2},.5)+(\tx,\ty)$) -- 
($\ts*( 0.16278934628247, -0.24650808451821)+\ts*.10*(-.5,-{sqrt(3)/2})+(\tx,\ty)$) node [darkblue,midway, above,rotate=-30] {$\gamma$}; 
\draw[thick,draw=darkblue,fill=darkblue] 
($\ts*( 0.16278934628247, -0.24650808451821)-\ts*.11*(.5,{sqrt(3)/2})+(\tx,\ty)$) -- 
($\ts*( 0.16278934628247, -0.24650808451821)-\ts*.09*(.5,{sqrt(3)/2})+(\tx,\ty)$);

\node[anchor=east] at ($\ts*(  -.02+0.16278934628247, -0.24650808451821)+(\tx,\ty)$) {$\color{darkblue}w$};


\draw[thick] (T1) -- (T2) -- (T3) -- cycle;
\end{tikzpicture}

%% file: appendix_b.tex
\section{One-Dimensional Quadrature Points and Weights for Approach 2}
\label{app:b}

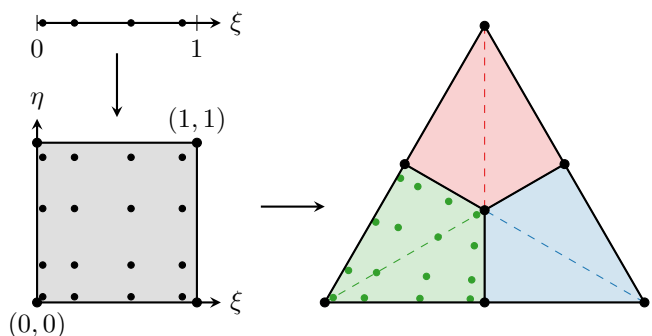
\begin{figure}
\centering
\input{mapping_ieee.tex}
\vspace{-1em}
\caption{Pictorial representation of Approach 2 to map one of the three quadrilaterals that form the original triangle.}
\label{fig:approach_2}
\end{figure}

Table~\ref{tab:approach_2} provides the one-dimensional points and weights for Approach 2.  Fig.~\ref{fig:approach_2} shows a pictorial representation of Approach 2 to subdivide a triangle into three quadrilaterals (more details in~\cite{freno_quad}).


\begin{table}[htbp]
\centering
\begin{tabular}[t]{ c c c}
\toprule
$n'$ & $w'$ & $\xi$ \\
\midrule
$1$ & $1.000000000000000$ & $0.500000000000000$ \\ \midrule
$2$ & $0.416878477229995$ & $0.158583759535360$ \\
    & $0.583121522770005$ & $0.744081339598618$ \\ \midrule
$3$ & $0.189997117971354$ & $0.068273669149223$ \\
    & $0.460255434822571$ & $0.408489594837560$ \\
    & $0.349747447206075$ & $0.854955900106044$ \\ \midrule
$4$ & $0.100882575161292$ & $0.035428798606880$ \\
    & $0.295788771158858$ & $0.234117483281889$ \\
    & $0.377297249516236$ & $0.587879558540673$ \\
    & $0.226031404163614$ & $0.908595817252184$ \\ \midrule
$5$ & $0.057875593510608$ & $0.020052668459088$ \\
    & $0.188381905418622$ & $0.140436646107533$ \\
    & $0.297642254345381$ & $0.389571975520278$ \\
    & $0.296296561222160$ & $0.698080385501447$ \\
    & $0.159803685503229$ & $0.936097423233341$ \\ \midrule
$6$ & $0.036467705409933$ & $0.012544980340007$ \\
    & $0.125413328436587$ & $0.090649292816740$ \\
    & $0.220348540816832$ & $0.265388636715017$ \\
    & $0.267514284509112$ & $0.515197424177348$ \\
    & $0.232620249111963$ & $0.772638772660649$ \\
    & $0.117635891715573$ & $0.953295709799319$ \\
\bottomrule
\end{tabular}
\caption{1D quadrature points and weights for Approach~2.}
\label{tab:approach_2}
\end{table}

\FloatBarrier

%% file: mapping_ieee.tex
\begin{tikzpicture}[scale=0.85]

\definecolor{lightgreen}{RGB}{173.4,217,170.6}
\definecolor{darkgreen} {RGB}{ 51  ,160, 44}
\definecolor{darkblue}  {RGB}{ 31  ,120,180}
\definecolor{lightblue} {RGB}{165.4,201,225}
\definecolor{darkred}   {RGB}{227  ,26,28}
\definecolor{lightred} {RGB}{243.8,163.4,164.2}

\def\ls{2.5}; 
\def\ss{2.5}; 
\def\ts{5}; 

\def\lx{+\ls*0-12}
\def\sx{+\ss*0-12}
\def\tx{+\ts*.5-7.5}

\def\ly{\ss*1.75}
\def\sy{\ss*0}
\def\ty{{\ts*sqrt(3)/6}}

\coordinate (A) at (-0.5,{-sqrt(3)/6});
\coordinate (B) at ( 0.5,{-sqrt(3)/6});
\coordinate (C) at ( 0  , {sqrt(1/3)});
\coordinate (D) at ($0.5*(A)+0.5*(B)$);
\coordinate (E) at ($0.5*(B)+0.5*(C)$);
\coordinate (F) at ($0.5*(C)+0.5*(A)$);
\coordinate (O) at (0,0);

\coordinate (T1) at ($\ts*(A)+(\tx,\ty)$);
\coordinate (T2) at ($\ts*(B)+(\tx,\ty)$);
\coordinate (T3) at ($\ts*(C)+(\tx,\ty)$);
\coordinate (T4) at ($\ts*(D)+(\tx,\ty)$);
\coordinate (T5) at ($\ts*(E)+(\tx,\ty)$);
\coordinate (T6) at ($\ts*(F)+(\tx,\ty)$);
\coordinate (TO) at ($\ts*(O)+(\tx,\ty)$);

\draw[fill=lightgreen,fill opacity=.5] (T1) -- (T4) -- (TO) -- (T6) -- cycle;
\draw[fill=lightblue, fill opacity=.5] (T2) -- (T5) -- (TO) -- (T4) -- cycle;
\draw[fill=lightred,  fill opacity=.5] (T3) -- (T6) -- (TO) -- (T5) -- cycle;

\draw[darkred,dashed] ($\ts*(C)+(\tx,\ty)$) -- ($\ts*(O)+(\tx,\ty)$);
\draw[darkgreen,dashed] ($\ts*(A)+(\tx,\ty)$) -- ($\ts*(O)+(\tx,\ty)$);
\draw[darkblue,dashed] ($\ts*(B)+(\tx,\ty)$) -- ($\ts*(O)+(\tx,\ty)$);

\draw[draw=darkgreen,fill=darkgreen] ($(T1)+\ts*(0.026257799012478, 0.015159947328181 )$) circle (.05);
\draw[draw=darkgreen,fill=darkgreen] ($(T1)+\ts*(0.074170144832526, 0.100178635875917 )$) circle (.05);
\draw[draw=darkgreen,fill=darkgreen] ($(T1)+\ts*(0.159477322317448, 0.251553072450514 )$) circle (.05);
\draw[draw=darkgreen,fill=darkgreen] ($(T1)+\ts*(0.236815739060366, 0.388787237326027 )$) circle (.05);
\draw[draw=darkgreen,fill=darkgreen] ($(T1)+\ts*(0.123842316001278, 0.014143911689380 )$) circle (.05);
\draw[draw=darkgreen,fill=darkgreen] ($(T1)+\ts*(0.161885363466855, 0.093464558175449 )$) circle (.05);
\draw[draw=darkgreen,fill=darkgreen] ($(T1)+\ts*(0.229620410596510, 0.234693720559197 )$) circle (.05);
\draw[draw=darkgreen,fill=darkgreen] ($(T1)+\ts*(0.291028154440107, 0.362730307147916 )$) circle (.05);
\draw[draw=darkgreen,fill=darkgreen] ($(T1)+\ts*(0.297590012300897, 0.012334876229172 )$) circle (.05);
\draw[draw=darkgreen,fill=darkgreen] ($(T1)+\ts*(0.318060929411206, 0.081510248524392 )$) circle (.05);
\draw[draw=darkgreen,fill=darkgreen] ($(T1)+\ts*(0.354509075068010, 0.204675909920681 )$) circle (.05);
\draw[draw=darkgreen,fill=darkgreen] ($(T1)+\ts*(0.387552506598853, 0.316336353160253 )$) circle (.05);
\draw[draw=darkgreen,fill=darkgreen] ($(T1)+\ts*(0.455107493721692, 0.010694827379250 )$) circle (.05);
\draw[draw=darkgreen,fill=darkgreen] ($(T1)+\ts*(0.459647737932681, 0.070672621387675 )$) circle (.05);
\draw[draw=darkgreen,fill=darkgreen] ($(T1)+\ts*(0.467731571276731, 0.177462139434817 )$) circle (.05);
\draw[draw=darkgreen,fill=darkgreen] ($(T1)+\ts*(0.475060273157097, 0.274276176588547 )$) circle (.05);
\draw[thick] ($\ts*(O)+(\tx,\ty)$) -- ($\ts*(D)+(\tx,\ty)$);
\draw[thick] ($\ts*(O)+(\tx,\ty)$) -- ($\ts*(E)+(\tx,\ty)$);
\draw[thick] ($\ts*(O)+(\tx,\ty)$) -- ($\ts*(F)+(\tx,\ty)$);

\draw[thick] (T1) -- (T2) -- (T3) -- cycle;

\draw[fill=black] (T1) circle (.07);
\draw[fill=black] (T2) circle (.07);
\draw[fill=black] (T3) circle (.07);
\draw[fill=black] (TO) circle (.07);
\draw[fill=black] (T4) circle (.07);
\draw[fill=black] (T5) circle (.07);
\draw[fill=black] (T6) circle (.07);

\coordinate (S1) at (0*\ss+\sx,0*\ss+\sy);
\coordinate (S2) at (1*\ss+\sx,0*\ss+\sy);
\coordinate (S4) at (0*\ss+\sx,1*\ss+\sy);
\coordinate (S3) at (1*\ss+\sx,1*\ss+\sy);
\draw[thick,fill=lightgray,fill opacity=.5] (S1) -- (S2) -- (S3) -- (S4) -- cycle;

\draw[fill=black] ($(S1)+\ss*(0.035428798606880, 0.035428798606880)$) circle (.05);
\draw[fill=black] ($(S1)+\ss*(0.035428798606880, 0.234117483281889)$) circle (.05);
\draw[fill=black] ($(S1)+\ss*(0.035428798606880, 0.587879558540673)$) circle (.05);
\draw[fill=black] ($(S1)+\ss*(0.035428798606880, 0.908595817252184)$) circle (.05);
\draw[fill=black] ($(S1)+\ss*(0.234117483281889, 0.035428798606880)$) circle (.05);
\draw[fill=black] ($(S1)+\ss*(0.234117483281889, 0.234117483281889)$) circle (.05);
\draw[fill=black] ($(S1)+\ss*(0.234117483281889, 0.587879558540673)$) circle (.05);
\draw[fill=black] ($(S1)+\ss*(0.234117483281889, 0.908595817252184)$) circle (.05);
\draw[fill=black] ($(S1)+\ss*(0.587879558540673, 0.035428798606880)$) circle (.05);
\draw[fill=black] ($(S1)+\ss*(0.587879558540673, 0.234117483281889)$) circle (.05);
\draw[fill=black] ($(S1)+\ss*(0.587879558540673, 0.587879558540673)$) circle (.05);
\draw[fill=black] ($(S1)+\ss*(0.587879558540673, 0.908595817252184)$) circle (.05);
\draw[fill=black] ($(S1)+\ss*(0.908595817252184, 0.035428798606880)$) circle (.05);
\draw[fill=black] ($(S1)+\ss*(0.908595817252184, 0.234117483281889)$) circle (.05);
\draw[fill=black] ($(S1)+\ss*(0.908595817252184, 0.587879558540673)$) circle (.05);
\draw[fill=black] ($(S1)+\ss*(0.908595817252184, 0.908595817252184)$) circle (.05);

\draw[fill=black] (S1) circle (.07);
\draw[fill=black] (S2) circle (.07);
\draw[fill=black] (S3) circle (.07);
\draw[fill=black] (S4) circle (.07);

\node[anchor=north] at (S1) {$(0,0)$};
\node[above] at (S3) {$(1,1)$};


\coordinate (L1) at (0*\ls+\lx,0*\ls+\ly);
\coordinate (L2) at (1*\ls+\lx,0*\ls+\ly);
\draw[thick] (L1) -- (L2);

\draw[fill=black] ($(L1)+\ls*(0.035428798606880, 0.0)$) circle (.05);
\draw[fill=black] ($(L1)+\ls*(0.234117483281889, 0.0)$) circle (.05);
\draw[fill=black] ($(L1)+\ls*(0.587879558540673, 0.0)$) circle (.05);
\draw[fill=black] ($(L1)+\ls*(0.908595817252184, 0.0)$) circle (.05);

\draw[fill=black] ($(L1)+\ls*(0.0,0.05)$) -- ($(L1)-\ls*(0.0,0.05)$);
\draw[fill=black] ($(L2)+\ls*(0.0,0.05)$) -- ($(L2)-\ls*(0.0,0.05)$);

\node[below = .1 of L1] {$0$};
\node[below = .1 of L2] {$1$};

\node[right] (xi1)  at ($(L2)+0.15*(L2)-0.15*(L1)$) {$\xi$};
\draw[thick,->,>=stealth] (L2) -- (xi1);

\node[above] (eta1) at ($(S4)+0.15*(S4)-0.15*(S1)$) {$\eta$};
\node[right] (xi1)  at ($(S2)+0.15*(S2)-0.15*(S1)$) {$\xi$};
\draw[thick,->,>=stealth] (S4) -- (eta1);
\draw[thick,->,>=stealth] (S2) -- (xi1);


\draw[thick,->,>=stealth] (-8.5,1.5) -- (-7.5,1.5);

\draw[thick,->,>=stealth] (0.5*\ls+\lx,3.9) -- (0.5*\ls+\lx,2.9);


\end{tikzpicture}

%% file: em_manuscript.bbl
\begin{thebibliography}{10}
\expandafter\ifx\csname url\endcsname\relax
  \def\url#1{\texttt{#1}}\fi
\expandafter\ifx\csname urlprefix\endcsname\relax\def\urlprefix{URL }\fi
\expandafter\ifx\csname href\endcsname\relax
  \def\href#1#2{#2} \def\path#1{#1}\fi

\bibitem{graglia_1993}
R.~D. Graglia, On the numerical integration of the linear shape functions times
  the {3-D} {Green's} function or its gradient on a plane triangle, {IEEE}
  Transactions on Antennas and Propagation 41~(10) (1993) 1448--1455.
\newblock \href {https://doi.org/10.1109/8.247786}
  {\path{doi:10.1109/8.247786}}.

\bibitem{wilton_1984}
D.~{Wilton}, S.~{Rao}, A.~{Glisson}, D.~{Schaubert}, O.~{Al-Bundak},
  C.~{Butler}, Potential integrals for uniform and linear source distributions
  on polygonal and polyhedral domains, {IEEE} Transactions on Antennas and
  Propagation 32~(3) (1984) 276--281.
\newblock \href {https://doi.org/10.1109/TAP.1984.1143304}
  {\path{doi:10.1109/TAP.1984.1143304}}.

\bibitem{rao_1982}
S.~{Rao}, D.~{Wilton}, A.~{Glisson}, Electromagnetic scattering by surfaces of
  arbitrary shape, {IEEE} Transactions on Antennas and Propagation 30~(3)
  (1982) 409--418.
\newblock \href {https://doi.org/10.1109/TAP.1982.1142818}
  {\path{doi:10.1109/TAP.1982.1142818}}.

\bibitem{khayat_2005}
M.~A. Khayat, D.~R. Wilton, Numerical evaluation of singular and near-singular
  potential integrals, {IEEE} Transactions on Antennas and Propagation 53~(10)
  (2005) 3180--3190.
\newblock \href {https://doi.org/10.1109/TAP.2005.856342}
  {\path{doi:10.1109/TAP.2005.856342}}.

\bibitem{fink_2008}
P.~W. Fink, D.~R. Wilton, M.~A. Khayat, Simple and efficient numerical
  evaluation of near-hypersingular integrals, {IEEE} Antennas and Wireless
  Propagation Letters 7 (2008) 469--472.
\newblock \href {https://doi.org/10.1109/LAWP.2008.2000788}
  {\path{doi:10.1109/LAWP.2008.2000788}}.

\bibitem{khayat_2008}
M.~A. Khayat, D.~R. Wilton, P.~W. Fink, An improved transformation and
  optimized sampling scheme for the numerical evaluation of singular and
  near-singular potentials, {IEEE} Antennas and Wireless Propagation Letters 7
  (2008) 377--380.
\newblock \href {https://doi.org/10.1109/LAWP.2008.928461}
  {\path{doi:10.1109/LAWP.2008.928461}}.

\bibitem{vipiana_2011}
F.~Vipiana, D.~R. Wilton, Optimized numerical evaluation of singular and
  near-singular potential integrals involving junction basis functions, {IEEE}
  Transactions on Antennas and Propagation 59~(1) (2011) 162--171.
\newblock \href {https://doi.org/10.1109/TAP.2010.2090464}
  {\path{doi:10.1109/TAP.2010.2090464}}.

\bibitem{vipiana_2012}
F.~Vipiana, D.~R. Wilton, Numerical evaluation via singularity cancellation
  schemes of near-singular integrals involving the gradient of {Helmholtz}-type
  potentials, {IEEE} Transactions on Antennas and Propagation 61~(3) (2013)
  1255--1265.
\newblock \href {https://doi.org/10.1109/TAP.2012.2227922}
  {\path{doi:10.1109/TAP.2012.2227922}}.

\bibitem{botha_2013}
M.~M. Botha, A family of augmented {D}uffy transformations for near-singularity
  cancellation quadrature, {IEEE} Transactions on Antennas and Propagation
  61~(6) (2013) 3123--3134.
\newblock \href {https://doi.org/10.1109/TAP.2013.2252137}
  {\path{doi:10.1109/TAP.2013.2252137}}.

\bibitem{rivero_2019}
J.~{Rivero}, F.~{Vipiana}, D.~R. {Wilton}, W.~A. {Johnson}, Hybrid integration
  scheme for the evaluation of strongly singular and near-singular integrals in
  surface integral equations, {IEEE} Transactions on Antennas and Propagation
  (2019).
\newblock \href {https://doi.org/10.1109/TAP.2019.2920333}
  {\path{doi:10.1109/TAP.2019.2920333}}.

\bibitem{vipiana_2013}
F.~Vipiana, D.~R. Wilton, W.~A. Johnson, Advanced numerical schemes for the
  accurate evaluation of {4-D} reaction integrals in the method of moments,
  {IEEE} Transactions on Antennas and Propagation 61~(11) (2013) 5559--5566.
\newblock \href {https://doi.org/10.1109/TAP.2013.2277864}
  {\path{doi:10.1109/TAP.2013.2277864}}.

\bibitem{ma_1996}
J.~Ma, V.~Rokhlin, S.~Wandzura, Generalized {G}aussian quadrature rules for
  systems of arbitrary functions, {SIAM} Journal on Numerical Analysis 33~(3)
  (1996) 971--996.
\newblock \href {https://doi.org/10.1137/0733048} {\path{doi:10.1137/0733048}}.

\bibitem{duffy_1982}
M.~Duffy, Quadrature over a pyramid or cube of integrands with a singularity at
  a vertex, {SIAM} Journal on Numerical Analysis 19~(6) (1982) 1260--1262.
\newblock \href {https://doi.org/10.1137/0719090} {\path{doi:10.1137/0719090}}.

\bibitem{polimeridis_2013}
A.~G. {Polimeridis}, F.~{Vipiana}, J.~R. {Mosig}, D.~R. {Wilton}, {DIRECTFN}:
  Fully numerical algorithms for high precision computation of singular
  integrals in {Galerkin} {SIE} methods, {IEEE} Transactions on Antennas and
  Propagation 61~(6) (2013) 3112--3122.
\newblock \href {https://doi.org/10.1109/TAP.2013.2246854}
  {\path{doi:10.1109/TAP.2013.2246854}}.

\bibitem{wilton_2017}
D.~R. {Wilton}, F.~{Vipiana}, W.~A. {Johnson}, Evaluation of {4-D} reaction
  integrals in the method of moments: Coplanar element case, {IEEE}
  Transactions on Antennas and Propagation 65~(5) (2017) 2479--2493.
\newblock \href {https://doi.org/10.1109/TAP.2017.2677916}
  {\path{doi:10.1109/TAP.2017.2677916}}.

\bibitem{rivero_2019b}
J.~Rivero, F.~Vipiana, D.~R. Wilton, W.~A. Johnson, Evaluation of {4-D} reaction
  integrals via double application of the divergence theorem, {IEEE}
  Transactions on Antennas and Propagation 67~(2) (2019) 1131--1142.
\newblock \href {https://doi.org/10.1109/TAP.2018.2882589}
  {\path{doi:10.1109/TAP.2018.2882589}}.

\bibitem{ylaoijala_2003}
P.~Yl\"{a}-Oijala, M.~Taskinen, Calculation of {CFIE} impedance matrix elements
  with {RWG} and $\boldsymbol{n}\times\mathrm{RWG}$ functions, {IEEE}
  Transactions on Antennas and Propagation 51~(8) (2003) 1837--1846.
\newblock \href {https://doi.org/10.1109/TAP.2003.814745}
  {\path{doi:10.1109/TAP.2003.814745}}.

\bibitem{gurel_2005}
L.~G\"{u}rel, O.~Erg\"{u}l, Singularity of the magnetic-field integral equation
  and its extraction, {IEEE} Antennas and Wireless Propagation Letters 4 (2005)
  229--232.
\newblock \href {https://doi.org/10.1109/LAWP.2005.851103}
  {\path{doi:10.1109/LAWP.2005.851103}}.

\bibitem{ergul_2005}
O.~Erg\"{u}l, L.~G\"{u}rel, Improved testing of the magnetic-field integral
  equation, {IEEE} Microwave and Wireless Components Letters 15~(10) (2005)
  615--617.
\newblock \href {https://doi.org/10.1109/LMWC.2005.856697}
  {\path{doi:10.1109/LMWC.2005.856697}}.

\bibitem{polimeridis_2012}
A.~G. Polimeridis, I.~D. Koufogiannis, M.~Mattes, J.~R. Mosig, Considerations
  on double exponential-based cubatures for the computation of weakly singular
  {Galerkin} inner products, {IEEE} Transactions on Antennas and Propagation
  60~(5) (2012) 2579--2582.
\newblock \href {https://doi.org/10.1109/TAP.2012.2189708}
  {\path{doi:10.1109/TAP.2012.2189708}}.

\bibitem{tihon_2018}
D.~Tihon, C.~Craeye, All-analytical evaluation of the singular integrals
  involved in the method of moments, {IEEE} Transactions on Antennas and
  Propagation 66~(4) (2018) 1925--1936.
\newblock \href {https://doi.org/10.1109/TAP.2018.2803130}
  {\path{doi:10.1109/TAP.2018.2803130}}.

\bibitem{dunavant_1985}
D.~A. Dunavant, High degree efficient symmetrical {G}aussian quadrature rules
  for the triangle, International Journal for Numerical Methods in Engineering
  21~(6) (1985) 1129--1148.
\newblock \href {https://doi.org/10.1002/nme.1620210612}
  {\path{doi:10.1002/nme.1620210612}}.

\bibitem{mathematica}
{Wolfram Research, Inc.}, Mathematica.

\bibitem{freno_quad}
B.~A. Freno, W.~A. Johnson, B.~F. Zinser, S.~Campione, Symmetric triangle
  quadrature rules for arbitrary functions, Computers \& Mathematics with
  Applications (2020).
\newblock \href {https://doi.org/10.1016/j.camwa.2019.12.021}
  {\path{doi:10.1016/j.camwa.2019.12.021}}.

\bibitem{lyness_1975}
J.~N. Lyness, D.~Jespersen, Moderate degree symmetric quadrature rules for the
  triangle, {IMA} Journal of Applied Mathematics 15~(1) (1975) 19--32.
\newblock \href {https://doi.org/10.1093/imamat/15.1.19}
  {\path{doi:10.1093/imamat/15.1.19}}.

\bibitem{wandzura_2003}
S.~Wandzura, H.~Xiao, Symmetric quadrature rules on a triangle, Computers \&
  Mathematics with Applications 45~(12) (2003) 1829--1840.
\newblock \href {https://doi.org/10.1016/S0898-1221(03)90004-6}
  {\path{doi:10.1016/S0898-1221(03)90004-6}}.

\bibitem{papanicolopulos_2015}
S.-A. Papanicolopulos, Computation of moderate-degree fully-symmetric cubature
  rules on the triangle using symmetric polynomials and algebraic solving,
  Computers \& Mathematics with Applications 69~(7) (2015) 650--666.
\newblock \href {https://doi.org/10.1016/j.camwa.2015.02.014}
  {\path{doi:10.1016/j.camwa.2015.02.014}}.

\bibitem{wilton_2019}
D.~R. {Wilton}, J.~{Rivero}, W.~A. {Johnson}, F.~{Vipiana}, Evaluation of
  static potential integrals on triangular domains, IEEE Access 8 (2020)
  99806--99819.
\newblock \href {https://doi.org/10.1109/ACCESS.2020.2997287}
  {\path{doi:10.1109/ACCESS.2020.2997287}}.

\end{thebibliography}
